\RequirePackage{fix-cm}
\documentclass[smallextended]{svjour3}       
\smartqed  
\usepackage[sort&compress,numbers]{natbib}
\usepackage{graphicx}
\usepackage{hyperref}
\hypersetup{
    colorlinks=true,
    linkcolor=magenta,
    filecolor=blue,      
    urlcolor=blue,
    citecolor=blue,
    pdftitle={Sharelatex Example},
    bookmarks=true,
    }
\usepackage{breqn}
\usepackage[linesnumbered,vlined,ruled]{algorithm2e}
\usepackage{amsfonts}   
\usepackage{amsmath}
\usepackage{graphicx}
\usepackage{tcolorbox}

\usepackage{url}
\begin{document}

\title{ Cost and Reliability Aware Scheduling of Workflows Across Multiple Clouds with Security Constraints}

\titlerunning{Workflow Scheduling in Multiple Clouds with Security Constraints}        

\author{Atherve Tekawade    \and
        Suman Banerjee}


\institute{ Atherve Tekawade  \at
              Department of Computer Science and Engineering,\\               
              Indian Institute of Technology Jammu. \\              
              \email{2018uee0137@iitjammu.ac.in }         
           \and
          Suman Banerjee \at
             Department of Computer Science and Engineering,\\ 
             Indian Institute of Technology Kharagpur. \\
             \email{suman.banerjee@iitjammu.ac.in}
             }
             

\date{Received: date / Accepted: date}

\maketitle

\begin{abstract}
Many real-world scientific workflows can be represented by a Directed Acyclic Graph (DAG), where each node represents a task and a directed edge signifies a dependency between two tasks. Due to the increasing computational resource  requirements of these workflows, they are deployed on multi-cloud systems for execution. In this paper, we propose a scheduling algorithm that allocates resources to the tasks present in the workflow using an efficient list-scheduling approach based on the parameters cost, processing time, and reliability. Next, for a  given a task-resource mapping, we propose a cipher assignment algorithm that assigns security services to edges responsible for transferring data in time-optimal manner subject to a given security constraint. The proposed algorithms have been analyzed to understand their time and space requirements. We implement the proposed scheduling and cipher assignment algorithm and experimented with two real-world scientific  workflows namely Epigenomics and Cybershake. We compare the performance of the proposed scheduling algorithm with the state-of-art evolutionary methods. We observe that our method outperforms the state-of-art methods always in terms of cost and reliability, and is inferior in terms of makespan in some cases.
\end{abstract}

\section{Introduction} \label{Sec:Intro}
In recent times, \emph{cloud computing} has emerged as an alternative computing framework and become popular due to many features including `pay as you use' kind of billing strategy, virtualization, rapid elasticity, on-demand use, and so on \cite{mastelic2014cloud}. A scientific workflow is defined as a set of tasks where there are several dependencies among tasks \cite{zhou2019minimizing}. In recent times 
cloud infrastructure has been used extensively used for the execution of workflows and many scientific workflows from different domains such as \emph{Epigenomics} in bio-informatics, \emph{Cybershake} in earthquake engineering, etc. has been successfully deployed on commercial clouds \cite{zhou2019minimizing,farid2020scheduling}.  Recently, due to large-scale computational resource and diversity requirements, multiple cloud providers club together to form a larger infrastructure, and such a framework is known as a multi-cloud system \cite{tang2021reliability}. In many cases, they are managed by a third party. Each one of them provides its own set of Virtual Machines (VMs) and billing mechanisms.

\paragraph{\textbf{Motivation}} As mentioned in the literature, recently the resource requirements for the execution of these workflows have been increased extensively \cite{guo2018cost}. In such a situation, one solution is to use a multi-cloud system to execute such gigantic size workflows. A multi-cloud system is highly heterogeneous with its respective hardware, software, and network infrastructure \cite{miraftabzadeh2016efficient}. Hence, it is prone to failure. As in a workflow, there are dependencies among the tasks so the failure of one task may lead to the failure of the entire workflow. For  successful execution of the workflow, the cloud infrastructure must provide highly reliable computing services. Also in a multi-cloud system, VMs of two different cloud services may reside in two different locations and the network link connecting two servers where these two VMs resides may be accessible to the adversaries. It may so happen that the two tasks allocated to these VMs have dependencies. So some kind of security measures needs to be imposed before transmitting the data from one VM to the other. As a whole for the  successful execution of a workflow in a multi-cloud system both security and reliability become important issues.

\paragraph{\textbf{Related Work}} In the past few years, there has been an extensive study on multi-cloud systems \cite{li2015data,diaz2015supporting,DBLP:conf/comsnets/TekawadeB23}. Several scheduling strategies have been proposed for multi-cloud systems \cite{miraftabzadeh2016efficient,kang2018dynamic}. Recently, there are a few methodologies for scheduling workflows in the multi-cloud systems as well  \cite{al2019task,sooezi2015scheduling,li2018scheduling}. In general, in a scientific workflow, tasks have a very high level of data and control flow dependency with the precedence task(s). Recently, there are several studies that focus on the reliability issues for workflow execution  \cite{han2018checkpointing,wen2016cost,di2016toward,zhou2016cloud}. Guo et al. \cite{guo2018cost} studied the problem of scheduling workflows to minimize the  cost subject to deadline constraints using a heuristic approach which is a   hybridization of the Genetic Algorithm and the Particle Swarm Optimization. Tang et al. \cite{tang2021reliability} considered the problem of scheduling workflows to minimize makespan, cost and maximize reliability. There are a few studies that consider both the security and reliability of the workflow. Wen et al. \cite{wen2016cost} proposed an algorithm to find deployments for workflows that are reliable, less expensive and satisfy security requirements in federated clouds. Zhu et al. \cite{zhu2021task} looked at the problem of task scheduling subject to reliability and security constraints. Apart from this, workflows are also known to be scheduled on multi-processor platforms taking into account the makespan, energy consumption and reliability requirements. \cite{zhou2016cloud} -\cite{xie2017energy}. To the best of our knowledge, there are limited literature on security and reliability of workflow execution. In this paper, we focus on these two aspects for workflow scheduling.
\paragraph{\textbf{Our Contributions}} In this paper, we make the following contributions:
\begin{itemize}
\item We propose a system model to study workflow scheduling in a multi-cloud system considering communication costs and billing mechanisms of different cloud providers.
\item We integrate the notion of data confidentiality in our model by assuming that data is encrypted using cryptographic ciphers of varying strengths. We develop an optimal assignment of ciphers to data to minimize the total encryption and decryption overhead.
\item We also perform the reliability analysis for task execution considering that the failure of the tasks is modeled using Poisson Distribution.
\item We propose a list-based heuristic followed by an iterative local-search algorithm that assigns the best resource to a task based on the makespan, cost, and reliability.
\item The proposed solution approach has been implemented with real-world scientific workflows and compared with the state-of-art approaches.
\end{itemize}   

\paragraph{Organization of the Paper} The rest of the paper is organized as follows. Section \ref{Sec:SM} describes the system's model and problem formulation. The proposed list-based heuristic solution has been described in Section \ref{Sec:PA}. Section \ref{Sec:Experiments} contains the experimental evaluation of the proposed solution methodology. Finally, Section \ref{Sec:CFA} concludes our study and gives future research directions.

\section{Systems Model \& Problem Formulation} \label{Sec:SM}
In this section, we describe the system's model and describe our problem formally. For any positive integer $n$, $[n]$ denotes the set $\{1, 2, \ldots, n\}$.
\subsection{Tasks and Workflow}
A task is a job that needs to be executed on the multi-cloud system. A \emph{scientific workflow} consists of multiple interrelated tasks and is modeled as a \emph{directed acyclic graph} (DAG) $G(V, E)$, where $V(G)=\{v_i : i \in [n]\}$ is the set of $n$ tasks and $E(G) = \{e_h = (v_i, v_j) : v_i, v_j \in V(G)\}$ is the set of edges. Here an edge $(v_i, v_j)$ signifies a precedence relationship between tasks $v_i$ and $v_j$, i.e., the execution of $v_j$ can only be started after the execution of $v_i$ is finished and its output is transferred to $v_j$. The weight of the edge $(v_iv_j)$ is denoted by $w_{v_i, v_j}$ and signifies the amount of data to be transferred. The computational resource requirement of the task $v_i$ is denoted by $w_{v_i}$. Next, we present some definitions related to workflow.
\begin{definition} [Predecessor of a Task] \label{def1}
For any task $v_i$, $pred(v_i)$ denotes the set of immediate predecessor tasks of this task and defined as $pred(v_i) = \{v_j : (v_j, v_i) \in E \} $.
\end{definition}  
\begin{definition} [Successor of a Task] \label{def2}
For any task $v_i$, $succ(v_i)$ denotes the set of immediate successor tasks of this task and defined as  $succ(v_i) = \{v_j : (v_i, v_j) \in E \} $.
\end{definition}
\begin{definition} [Entry Task] \label{def3}
 It is a redundant task denoted by $v_{entry}$ having an outgoing edge with zero weight to every $v$ such that $pred(v) = \emptyset$. 
\end{definition}
\begin{definition} [Exit Task] \label{def4}
$v_{exit}$ denotes the exit task. It is a redundant node having an incoming edge with zero weight from every $v$ such that $succ(v) = \emptyset$. 
\end{definition}
For simplicity, we assume that $v_{entry} = v_1$ and $v_{exit} = v_n$.
\begin{definition} [Topological Level] \label{def5}
For a task $v$, $top\_level(v)$ denotes its topological level and defined by Equation No. \ref{eqn1}.
\begin{equation} \label{eqn1}
\scriptsize
top\_level(v) = \left\{ 
  \begin{array}{ c l }
   0 & \quad \textrm{if } v = v_1 \\
  \underset{u \in pred(v)}{max}   \{top\_level(u) + 1\} & \quad \textrm{otherwise}
  \end{array}
\right.
\end{equation}
\end{definition}
The topological level can be computed using the Breadth-first search algorithm (BFS) \cite{bundy1984breadth}.

\subsection{Multi-Cloud System}
Our model consists of $m$ different cloud providers, each providing their resources. These resources are offered in the form of Virtual Machines (henceforth mentioned as VMs) \cite{stillwell2012virtual}. Assume that the $k^{th}$ cloud provider offers a total of $m_k$ different VM types. Let $VM(k, p)$ denote the $p^{th}$ type VM offered by the $k^{th}$ cloud provider. A VM is characterized by its CPU, Disk, and Memory. We assume that VMs have sufficient memory to execute the
workflow tasks \cite{rodriguez2014deadline}. Let $w_{k, p}$ denote the processing capacity of $VM(k, p)$, which is proportional to the MIPS (amount of computation performed per second). The higher the processing capacity of a VM, the faster it executes a task. The amount of time it takes to execute $v_i$ on $VM(k, p)$ is denoted by $T_{exec}[v_i, VM(k, p)]$ and is given by Equation No. \ref{eqn2}. Additionally, we assume that a newly launched VM needs a specific initial boot time denoted by $T_{boot}[VM(k, p)]$.

\begin{equation} \label{eqn2}
T_{exec}[v_i, VM(k,p)] = \frac{w_{v_i}}{w_{k,p}}
\end{equation}
\subsection{Network}
The bandwidth between two VMs depends on multiple factors like their physical location, the cloud provider, etc. For simplicity, we assume VMs belonging to the same center are connected by a high-speed internal network. In contrast, those belonging to different centers are connected by slower external networks \cite{tang2021reliability}. Let $B_k$ and $B_{k, k'}$ denote the bandwidth of the communication links of the VMs within the $k$-th cloud and the bandwidth of the communication links connecting the $k$-th and $k^{'}$-th clouds, respectively. Communication time between two tasks $v_i$ and $v_j$ is denoted by $T_{comm}[v_i, v_j]$. This depends on the amount of data to be transferred and the clouds where the tasks are hosted. This can be computed using Equation No. \ref{eqn3}.

\begin{equation} \label{eqn3}
\scriptsize
T_{comm}[v_i, v_j] = \left\{ 
  \begin{array}{ c l }
  0 & \quad \textrm{if } \text{$v_i, v_j$ are scheduled on the same VM instance} \\
  \frac{w_{v_i, v_j}}{B_k} & \quad \textrm{else if } k' = k \\
    \frac{w_{v_i, v_j}}{B_{k, k'}} & \quad \textrm{otherwise}
  \end{array}
\right.
\end{equation}
\subsection{Resource Model and Timing Metrics}
Due to the flexibility of resource acquisition provided by cloud providers, a client can run any number of instances of a given VM \cite{rodriguez2014deadline}. Hence the maximum number of instances can be at most $\mathcal{N} \cdot n$, where $\mathcal{N} = \sum_{k=1}^{m} \sum_{p=1}^{m_k} 1$ assuming each task is run on every possible VM type. Let $R = \{vm_r : r \in [\mathcal{N} \cdot n]\}$ denote the set of VM instances to be leased (pool of resources), and $LST_{vm_r}$ and $LFT_{vm_r}$ denote the start and end times respectively for which $vm_r$ is leased. For task $v_i$, once all the data between $v_j$ and $v_j$, $\forall v_j \in pred(v_i)$ is received, the decryption starts happening. Then, execution of the task $v_i$ on its allocated VM starts happening, after which encryption of all data between $v_i$ and $v_j$, $\forall v_j \in succ(v_i)$ happens. Then finally, the data needs to be transferred. A VM must be kept on until a task has transferred data to all its successor nodes \cite{rodriguez2014deadline}. Hence the total processing time denoted by $PT_{v_i}$ includes the decryption, execution, encryption, and communication time as shown in Equation No. \ref{eqn4}. 

\begin{dmath} \label{eqn4}
\scriptsize
PT_{v_i} = \left\{ 
  \begin{array}{ c l }
  T_{exec}[v_i, VM(k,p)] \\ + \sum_{v_j \in succ(v_i)} (T_{enc}[v_i, v_j] + T_{comm}[v_i, v_j]) & \quad \textrm{if } v_i = v_1 \\ \\
  \sum_{v_j \in pred(v_i)} T_{dec}[v_j, v_i] + T_{exec}[v_i, VM(k,p)] & \quad \textrm{else if } v_i = v_n \\ \\
  \sum_{v_j \in pred(v_i)} T_{dec}[v_j, v_i] + T_{exec}[v_i, VM(k,p)] \\ + \sum_{v_j \in succ(v_i)} (T_{enc}[v_i, v_j] + T_{comm}[v_i, v_j]) & \quad \textrm{otherwise}
  \end{array}
\right.
\end{dmath}
where $v_i$ is assumed to be executed on VM of type $VM(k, p)$.
Let the start and finish times of task $v_i$ be denoted by $ST_{v_i}$ and $FT_{v_i}$, respectively. As shown in Equation No, a task can start only after all its predecessors are complete. \ref{eqn5}.

\begin{equation} \label{eqn5}
ST_{v_i} = \left\{ 
  \begin{array}{ c l }
  0 & \quad \textrm{if } v_i = v_1 \\
  \max_{v_j \in pred(v_i)} FT_{v_j} & \quad \textrm{otherwise}
  \end{array}
\right.
\end{equation}

The finish time is given by the sum of start and processing times, as shown in Equation No. \ref{eqn6}.

\begin{equation} \label{eqn6}
FT_{v_i} = ST_{v_i} + PT_{v_i}
\end{equation}

The makespan is the total time required to execute the workflow which happens when the last task finishes.

\begin{equation} \label{eqn7}
makespan = FT_{v_n}
\end{equation}

\subsection{Pricing Mechanisms}
In this study, we consider three popular cloud-providing services: Microsoft Azure (\emph{MA}), Amazon Web Services (\emph{AWS}), and Google Cloud Platform (\emph{GCP}). Each cloud provider charges the customer after a specified billing period $\tau$. Let $c_{k, p}$ denote the price for renting $VM(k,p)$ for a single billing period. Let the cost corresponding to renting VM $vm_r$ be denoted by $C_{vm_r}$. Below we discuss the pricing mechanisms used by different cloud providers \cite{tang2021reliability}.

\begin{itemize}
    \item \emph{MA} follows a fine-grained scheme where the customer is charged per minute of usage \emph{i.e.} $\tau = $ 1 min. The cost is given in Equation No. \ref{eqn8}.
    
    \begin{equation} \label{eqn8}
        C_{vm_r} = \lceil \frac{LFT_{vm_r} - LST_{vm_r}}{\tau} \rceil \cdot c_{k, p}
    \end{equation}
    
    \item \emph{AWS} follows a coarse-grained pricing mechanism where the customer per hour of usage. \emph{i.e.} $\tau = $ 1 hr. Equation No can obtain the cost. \ref{eqn8}.
    
    \item \emph{GCP} follows a hybrid pricing mechanism where the customer is charged for a minimum of ten minutes, after which per-minute billing is followed. The cost is formulated in Equation No. \ref{eqn9} where $C_{k,p}$ denotes the price for the first ten minutes and $\tau = $ 1 min.
    
    \begin{equation} \label{eqn9}
    	\scriptsize
        C_{vm_r} = C_{k, p} + \max(0, \lceil \frac{LFT_{vm_r} - LST_{vm_r} - 10 \cdot \tau}{\tau} \rceil) \cdot c_{k, p}
    \end{equation}

\end{itemize}
Table \ref{tab1} illustrates the prices of various VMs \cite{tang2021reliability}. Apart from this, each cloud provider has a specific pricing scheme associated with sending data out. Generally, ingress data is not charged \cite{ma}, \cite{aws}, \cite{gcp}. Let $c_{k, k'}$ denote the price per unit data for sending data between the $k$-th and $k^{'}$-th providers. This price depends on many factors: Location where the VMs are hosted, size of data, etc. Table \ref{tab2} shows the rates for transferring data as taken from the official websites \cite{ma}, \cite{aws}, \cite{gcp}. In the table, across centers refers to locations managed by the same cloud provider but located in different places. The prices vary depending on the location (US, Europe, Asia, etc.). For simplicity, we consider the median value across all locations. Across clouds refers to the transfers across different cloud providers over the external internet. Again this price depends on the location, so we consider the median value. The cost for transferring data between $v_i$ and $v_j$ is denoted by $C_{v_i, v_j}$ and is given by Equation No. \ref{eqn10}.

\begin{equation} \label{eqn10}
C_{v_i, v_j} = w_{v_i, v_j} \cdot c_{k, k'}
\end{equation}
where $v_i, v_j$ are assumed to be scheduled on the $k$-th and $k^{'}$-th clouds respectively. \\

As shown in Equation No, the total execution cost is the sum of costs associated with task execution on a VM and the cost of transferring data from one task to another. \ref{eqn11}.

\begin{equation} \label{eqn11}
cost = \sum_{vm_r \in R} C_{vm_r} + \sum_{i=1}^{n} \sum_{v_j \in succ(v_i)} C_{v_i, v_j} 
\end{equation}

\begin{table*}
		\centering
		\begin{tabular}{|c|c|c|c|c|c|c|} 
			\hline
			\multicolumn{2}{|c|}{\emph{MA}} & \multicolumn{2}{c|}{\emph{AWS}} & \multicolumn{3}{c|}{\emph{GCP}} \\
			\hline
			\textbf{\emph{VM}} & \textbf{\emph{Per Minute(\$)}} & \textbf{\emph{VM}} & \textbf{\emph{Per hour(\$)}} & \textbf{\emph{VM}} & \textbf{\emph{Ten Minutes(\$)}} & \textbf{\emph{Per Minute(\$)}}\\
			\hline
			\textbf{B2MS} & 0.0015 & \textbf{m1.small} & 0.06 & \textbf{n1-highcpu-2} & 0.014 & 0.0012 \\
			\hline
			\textbf{B4MS} & 0.003 & \textbf{m1.medium} & 0.12 & \textbf{n1-highcpu-4} & 0.025 & 0.0023 \\
			\hline
			\textbf{B8MS} & 0.006 & \textbf{m1.large} & 0.24 & \textbf{n1-highcpu-8} & 0.05 & 0.0047 \\
			\hline
			\textbf{B16MS} & 0.012 & \textbf{m1.xlarge} & 0.45 & \textbf{n1-highcpu-16} & 0.1 & 0.0093 \\
			\hline
		\end{tabular}
		\vspace{0.1 cm}
		\caption{VM costs for different cloud providers}
	\label{tab1}
\end{table*}

\begin{table*}
		\centering
		\begin{tabular}{|c|c|c|c|c|c|c|} 
			\hline
			\multicolumn{2}{|c|}{\emph{MA}} & \multicolumn{2}{c|}{\emph{AWS}} & \multicolumn{2}{c|}{\emph{GCP}} \\
			\hline
			\multicolumn{2}{|c|}{\textbf{Same center - Free}} & \multicolumn{2}{c|}{\textbf{Same center- Free}} & \multicolumn{2}{c|}{\textbf{Same center - Free}} \\
			\hline
			\multicolumn{2}{|c|}{\textbf{Across centers - \$0.08/GB}} & \multicolumn{2}{c|}{\textbf{Across centers - \$0.02/GB}} & \multicolumn{2}{c|}{\textbf{Across centers - \$0.05/GB}} \\
			\hline
			\multicolumn{6}{|c|}{\textbf{Across clouds}} \\
			\hline
			\textbf{\emph{Data size}} & \textbf{\emph{Per GB(\$)}} & \textbf{\emph{Data size}} & \textbf{\emph{Per GB(\$)}} & \textbf{\emph{Data size}} & \textbf{\emph{Per GB(\$)}}\\
			\hline
			Upto 100GB & Free & Upto 100GB & Free & 0-1TB & 0.19 \\
			\hline
			First 10 TB & 0.11 & First 10 TB & 0.09 & 1-10TB & 0.18 \\
			\hline
			Next 40TB & 0.075 & Next 40TB & 0.085 & 10TB+ & 0.15 \\
			\hline
			Next 100TB & 0.07 & Next 100TB & 0.07 & \multicolumn{2}{|c|}{-} \\ 
			\hline
			Next 350TB & 0.06 & Greater than 150 TB & 0.05 & \multicolumn{2}{|c|}{-} \\ 
			\hline
		\end{tabular}
		\vspace{0.1 cm}
		\caption{Data-Transfer costs for different cloud providers}
	\label{tab2}
\end{table*}

\begin{table}
\begin{tabular}{||c c c c c||} 
 \hline
 Level & Rounds & Plaintexts & Vul & Time ($\mu s/block$) \\ [0.5ex] 
 \hline\hline
 1 & 4 & $2^{29}$ & 98 & 3.08 \\ 
 \hline
 2 & 8 & $2^{61}$ & 67 & 3.58 \\
 \hline
 3 & 12 & $2^{94}$ & 34 & 4.15 \\
 \hline
 4 & 16 & $2^{118}$ & 10 & 4.63 \\
 \hline
 5 & 20 & $2^{128}$ & 0 & 5.21 \\ [1ex] 
 \hline
\end{tabular}
\vspace{0.1 cm}
\caption{\label{tab3} Security comparison of RC6 variants.}
\end{table}

\subsection{Security Mechanism}
Due to data dependencies between tasks, data needs to be transferred between tasks. To ensure secure communication, different security mechanisms can be employed to achieve the goal of  confidentiality. Similar to \cite{jiang2017design}, we focus on achieving data confidentiality using block ciphers. The notion of \emph{Security Level} is used to measure of the strength of a cipher. The security level is proportionally related to the number of rounds of encryption. The designer has to make a trade off between security and time: To ensure more security, we must go for a cipher with more encryption rounds. Table \ref{tab3} captures this trade off, where each row corresponds to one cipher with level $L_i$. The Plaintexts column represents the number of plaintexts needed for a successful cryptanalysis attack, and the Time column represents the time required to encrypt a block (128 bits) of data. The Vul column represents the vulnerability $V_i$ defined as the logarithm of the ratio of the maximum number of plaintexts required by the brute-force search, $PT_{bf}$, to the number of plaintexts required using a chosen cryptanalysis algorithm, $PT_{cc}(L_i)$ as shown in Equation No. \ref{eqn12} \cite{chandramouli2006battery}. The table illustrates the comparison of parameters for different variants of RC6, a widely used block cipher \cite{jiang2011optimization}. 

\begin{equation} \label{eqn12}
V_i = \log_2 \lfloor \frac{PT_{bf}}{PT_{cc}(L_i)} \rfloor
\end{equation}

The system vulnerability is defined as a weighted sum of the vulnerabilities of each data item, as shown in Equation No. \ref{eqn13} \cite{jiang2017design}.

\begin{equation} \label{eqn13}
V_{system} = \sum_{i=1}^{n} \sum_{v_j \in succ(v_i)} W_{v_i, v_j} \cdot V_{v_i, v_j}
\end{equation}
where $W_{v_i, v_j}$ and $V_{v_i, v_j}$ denote the weight and vulnerability of the data between $v_i$ and $v_j$ respectively. The maximum vulnerability $V_{\max}$ is got by setting $V_{v_i, v_j}$ to the cipher with the maximum vulnerability in Equation \ref{eqn13}.

The encryption time can be characterized by a linear function of data size and selected security level \cite{xie2007improving}. If the data between $v_i$ and $v_j$ is encrypted using level $L_i$ security, the time required to encrypt is denoted by $T_{enc}[v_i, v_j]$ given by Equation No. \ref{eqn14}.
\begin{dmath} \label{eqn14}
\scriptsize
 T_{enc}[v_i, v_j] =  \left\{ 
  \begin{array}{ c l }
   0 & \quad \textrm{if } \text{$v_i, v_j$ are scheduled on the same VM instance} \\
   t_i \cdot \frac{w_{v_i, v_j}}{B} \cdot \frac{1}{w_{k, p}} & \quad \textrm{otherwise}
  \end{array}
\right.
\end{dmath}

where $t_i$ denotes the amount of time required to encrypt one block of data using cipher of level $L_i$, $B$ denotes the block size (128 bits), and the $v_i$ is assumed to be executed on $VM(k,p)$. The decryption overhead is similar to that of encryption \cite{jiang2017design} and is denoted by $T_{dec}[v_i, v_j]$, the only difference is that it is executed on the VM executing $v_j$.
\subsection{Reliability Analysis}
Reliability is defined as the probability of a failure-free execution of the workflow. The occurrence of a failure is modeled as a Poisson Distribution \cite{farid2020scheduling}. Let $\lambda_{k, p}$ denote the parameter of the distribution for $VM(k, p)$. Similarly, let $\lambda_{k}$ and $\lambda_{k', k}$ denote the parameters for the communication links for the $k$-th cloud and for those between the $k^{'}$-th and the $k$-th cloud, respectively. The probability that the data will be transferred successfully between $v_i$ and $v_j$ is denoted by $\mathcal{R}_{v_i, v_j}$ and is given by Equation No. \ref{eqn15}. 

\begin{dmath} \label{eqn15}
\scriptsize
\mathcal{R}_{v_i, v_j} = \left\{ 
  \begin{array}{ c l }
   1 & \quad \textrm{if } \text{$v_i, v_j$ are scheduled on the same VM instance} \\
   e^{-\lambda_{k} \cdot T_{comm}[v_i, v_j]} & \quad \textrm{if } k' = k \\
   e^{-\lambda_{k', k} \cdot T_{comm}[v_i, v_j]} & \quad \textrm{otherwise}
  \end{array}
\right.
\end{dmath}
where $v_i, v_j$ are assumed to be scheduled on the $k^{'}$-th and $k$-{th} clouds respectively. \\

The probability that the VM instance $vm_r$ will execute successfully is denoted by $\mathcal{R}_{vm_r}$ and is given in Equation No. \ref{eqn16}.

\begin{equation} \label{eqn16}
\mathcal{R}_{vm_r} = e^{-\lambda_{k, p} \cdot (LFT_{vm_r} - LST_{vm_r})}
\end{equation}

The workflow will execute successfully if all the data is transferred and all the tasks are executed successfully. Assuming the failures are independent, Equation No  \ref{eqn17} gives the workflow's reliability.

\begin{equation} \label{eqn17}
reliability = \prod_{i=1}^{n} \prod_{v_j \in pred(v_i)} \mathcal{R}_{v_j, v_i} \cdot \prod_{vm_r \in R} \mathcal{R}_{vm_r}
\end{equation}

\subsection{Problem Definition}
A schedule $\mathcal{S} = (R, \mathcal{M}, \mathcal{C}, makespan, cost, reliability)$ is defined as a tuple consisting of: A pool of resources ($R$), a task to resource mapping ($\mathcal{M}$), and an edge to cipher mapping ($\mathcal{C}$). $R$ is a mapping from each resource to the times it is leased for, consisting of tuples of the form $(vm_r, LST_{vm_r}, LFT_{vm_r})$. $\mathcal{M}$ is a mapping consisting of tuples of the form $(v_i, vm_{u_i}, ST_{v_i}, FT_{v_i})$, where task $v_i$ is allocated to resource $vm_{u_i}$.  $\mathcal{C}$ is a mapping consisting of tuples of the form $(v_i, v_j, C_{v_i, v_j})$, where $v_j \in succ(v_i)$ and $C_{v_i, v_j}$ is the cipher used to encrypt the data between $v_i$ and $v_j$. The problem we address in this paper is that of minimizing the total makespan and cost and maximizing reliability subject to the security constraints stated below. \\

\textbf{Minimize} $\left\{ 
  \begin{array}{ c l }
   	 \textit{makespan} \\
   	 \textit{cost}
  \end{array}
\right.$
\textbf{Maximize} $\left\{ 
  \begin{array}{ c l }
  	 \textit{reliability}
  \end{array}
\right.$ \\ \\
\textbf{Subject To} $\left\{ 
  \begin{array}{ c l }
  	 V_{system} \leq UV_{req} \\
  	 V_{v_i, v_j} \leq  UV_{v_i, v_j}
  \end{array}
\right.$ \\ \\
where $UV_{req}$ and $UV_{v_i, v_j}$ denote the upper bounds on the system and data item between $v_i$ and $v_j$ vulnerabilities, respectively.  The trade-off is between more security and time-overhead leading to more makespan, cost and lower reliability. Before presenting our algorithm, we first present the pseudocode to convert a task to resource mapping to a schedule in Algorithm 1 similar to the one in \cite{guo2018cost} and is described as follows. We initialize the resources leased till now $R_{curr}$ and task to resource mapping $\mathcal{M}$ to empty. If a task has predecessors, it can start only after they finish, as illustrated in Lines No. 9, 10. For each task, we find the decryption, encryption, transfer times, transfer costs, and reliability associated with transferring data using the $process\_task()$ routine in Algorithm 2. The routine simply computes the encryption, decryption time using Equation No. \ref{eqn14}, transfer time using Equation No. \ref{eqn3}, transfer cost using Equation No. \ref{eqn10} and reliability using Equation No. \ref{eqn15}. The processing time is computed in Line No. 13 using Equation No. \ref{eqn4}. If the resource is already hosted, we know that the task can start only after the current lease finish time $LFT_{vm_{u_i}}$ as shown in Line No. 15, 16. Otherwise, we first launch the corresponding VM instance with lease start time $LST_{vm_{u_i}}$ set as shown in Line No. 19 and task start time after booting as shown in Line No. 18. Finally, the finish time of the task is calculated by adding the processing time to the start time as shown in Line No. 21. In Line No. 22, the lease finish time updated to the finish time of the task. After processing all the tasks, the makespan is given by the finish time of the last task as shown in Line No. 24. Lines 25 to 30 compute the cost and reliability associated with executing tasks.
  
\begin{algorithm}[t]
\caption{Particle to Schedule Mapping}
	\KwIn{$G(V, E)$, Multi-cloud system parameters, Resource Pool ($R$), Security Cipher Table similar to Table \ref{tab3} ($cipher\_tab$), Data to cipher mapping ($\mathcal{C}$), Task to resource mapping ($X$)}
    \KwOut{$\mathcal{S}$}
    $\mathcal{M}, R_{curr} \longleftarrow \emptyset$\;
    $cost \longleftarrow 0$\;
    $reliability \longleftarrow 1$\;
    \For{$v_i \in V$}{
        $vm_{u_i} \longleftarrow R[X[v_i]]$\;
        $VM(k, p) \longleftarrow type(vm_{u_i})$\;
        $exec\_time \longleftarrow \frac{w(v_i)}{w(VM(k, p))}$\;
        $ST_{v_i} \longleftarrow 0$\;
        \For{$v_j \in pred(v_i)$}{
            $ST_{v_i} \longleftarrow \max(ST_{v_i}, FT_{v_j})$\;
        }
        $dec\_time, transfer\_time, transfer\_cost, enc\_time, rel \longleftarrow process\_task(v_i)$\;
        $reliability \longleftarrow reliability \cdot rel$\;
        $PT_{v_i} \longleftarrow dec\_time + exec\_time + transfer\_time + enc\_time$\;
        $cost \longleftarrow cost + transfer\_cost$\;
        \If{$vm_{u_i} \in R_{curr}$}{
            $ST_{v_i} \longleftarrow \max(ST_{v_i}, LFT_{vm_{u_i}})$\;
        }
        \Else{
            $ST_{v_i} \longleftarrow \max(ST_{v_i}, T_{boot}[VM(k,p)])$\;
            $LST_{vm_{u_i}} \longleftarrow ST_{v_i} - T_{boot}[VM(k,p)]$\;
            $R_{curr} \longleftarrow R_{curr} \cup \{vm_{u_i}\}$\;
        }
        $FT_{v_i} \longleftarrow ST_{v_i} + PT_{v_i}$\;
        $LFT_{vm_{u_i}} \longleftarrow FT_{v_i}$\;
        $\mathcal{M} \longleftarrow \mathcal{M} \cup \{(v_i, vm_{u_i}, ST_{v_i}, FT_{v_i})\}$\;
    }
    $makespan \longleftarrow FT_{v_n}$\;
    \For{$vm_r \in R_{curr}$}{
    	$VM(k, p) \longleftarrow type(vm_r)$\;
        Compute $cost_{vm_r}$ corresponding to lease period $LFT_{vm_r}-LST_{vm_r}$ using Equation No. \ref{eqn8}, \ref{eqn9}\;
        $cost \longleftarrow cost + cost_{vm_r}$\;
        Compute $\mathcal{R}_{vm_r}$ corresponding to lease period $LFT_{vm_r}-LST_{vm_r}$ using Equation No. \ref{eqn16}\;
        $reliability \longleftarrow reliability \cdot \mathcal{R}_{vm_r}$\;
    }
    $\mathcal{S} \longleftarrow (R, \mathcal{M}, \mathcal{C}, makespan, cost, reliability)$\;
	\KwRet $\mathcal{S}$\;
	
	\label{Algo:1}
\end{algorithm}

\begin{algorithm}[t]
\caption{$process\_task$}
	\KwIn{$G(V, E)$, Multi-cloud system parameters, Resource Pool ($R$), Security Cipher Table similar to Table \ref{tab3} ($cipher\_tab$), Data to cipher mapping ($\mathcal{C}$), Task to resource mapping ($X$), $v_i$}
    \KwOut{$dec\_time, transfer\_time, transfer\_cost,$ \\ $enc\_time, rel$}
    $vm_{u_i} \longleftarrow R[X[v_i]]$\;
        $VM(k, p) \longleftarrow type(vm_{u_i})$\;
        $dec\_time, transfer\_time, transfer\_cost, enc\_time \longleftarrow 0$\;
        $rel \longleftarrow 1$\;
        \For{$v_j \in pred(v_i)$}{
        		$vm_{u_j} \longleftarrow R[X[v_j]]$\;
            \If{$u_i \neq u_j$}{
            	Find $cipher$ used to encrypt data between $v_j$ and $v_i$ from $\mathcal{C}$\;
            Find $T_{dec}[v_j, v_i]$ encrypted using $cipher$ and executed on $VM(k, p)$ from Equation No. \ref{eqn14}\;
            $dec\_time \longleftarrow dec\_time + T_{dec}[v_j, v_i]$\;
            }
           	}
\For{$v_j \in succ(v_i)$}{
			$vm_{u_j} \longleftarrow R[X[v_j]]$\;
			$VM(k', p') \longleftarrow type(vm_{u_i})$\;
            \If{$u_i \neq u_j$}{
            	Find $cipher$ used to encrypt data between $v_i$ and $v_j$ from $\mathcal{C}$\;
            Find $T_{enc}[v_i, v_j]$ encrypted using $cipher$ and executed on $VM(k, p)$ from Equation No. \ref{eqn14}\;
                \If{$k' \neq k$}{
                    $transfer\_time \longleftarrow transfer\_time + \frac{w(v_i, v_j)}{B_{k, k'}}$\;
                     Find pricing $c_{k, k'}$ according to Table \ref{tab2}\;
                     $transfer\_cost \longleftarrow transfer\_cost + c_{k, k'} \cdot w(v_i, v_j)$\;
                     $rel \longleftarrow rel \cdot e^{\lambda_{k, k'} \cdot \frac{w(v_i, v_j)}{B_{k, k'}}}$\;
                }
                \Else{
                    $transfer\_time \longleftarrow transfer\_time + \frac{w(v_i, v_j)}{B_{k}}$\;
                    $rel \longleftarrow rel \cdot e^{\lambda_{k} \cdot \frac{w(v_i, v_j)}{B_{k}}}$\;
                }
                $enc\_time \longleftarrow enc\_time + T_{enc}[v_i, v_j]$\;
            }
        }
        \KwRet $dec\_time, transfer\_time, transfer\_cost, enc\_time, rel$\;
	\label{Algo:2}
\end{algorithm}

\section{Proposed Algorithm} \label{Sec:PA}
Our solution methodology is divided into two parts: assigning each task to a VM and then the appropriate ciphers. We first describe the process employed to assign ciphers to data because this applies to any task to resource allocation.

\subsection{Cipher Assignment}
Given a particular task to resource mapping, we wish to assign the ciphers to data so that the total encryption and decryption overhead is minimized subject to the security constraints as stated below. \\

\textbf{Minimize} $\left\{ 
  \begin{array}{ c l }
  	 \sum_{i=1}^{n} \sum_{v_j \in succ(v_i)} T_{enc}[v_i, v_j] + T_{dec}[v_i, v_j]
  \end{array}
\right.$ \\ \\
\textbf{Subject To} $\left\{ 
  \begin{array}{ c l }
  	 V_{system} \leq UV_{req} \\
  	 V_{v_i, v_j} \leq  UV_{v_i, v_j}
  \end{array}
\right.$ \\ \\

We use a Dynamic Programming (DP) strategy to solve this problem as presented in Algorithm 3 and described as follows. The DP entry $dp[h][Vul]$ describes the least time to encrypt and decrypt the first $h$ edges and the corresponding cipher for the $h-${th} edge, as shown in Line No. 21 for the security constraint $V_{system} \leq Vul$. We begin by finding the VMs that the tasks $v_i, v_j$ are executed on in Line No 4. Then assignment of all possible ciphers to the $h-${th} edge is done in Line No. 8. If the constraints in Line No. 9 are satisfied, we calculate the encryption and decryption time in Line No. 10. If this is the first edge ($h = 1$), we find the least encryption time as in Line No. 13. If not, we find the remaining vulnerability $V_{remaining}$ in Line No. 16 that the first $h-1$ edges should satisfy using Equation No. \ref{eqn13}. The total time will be given by the best time to encrypt the $h-1$ edges and the time to encrypt the $h-${th} edge, which is calculated in Line No. 17. The total minimum time is updated in Line No. 19. The overall recurrence relation is shown in Equation No. \ref{recc}. Finally, we begin assigning the ciphers to each edge in Line No. 23- 27. We start adding the ciphers in a reverse manner \emph{i.e.} first the last edge will be assigned the cipher, then the second last, and so on. As mentioned earlier, the DP entry $dp[\lvert E \rvert][UV_{req}]$ stores the total minimum time and the cipher corresponding to the last edge for the constraint of $UV_{req}$. Hence, we start with $Vul = UV_{req}$ as shown in Line No. 22. The corresponding cipher is got from the dp table entry $dp[h][Vul]$ in Line No. 25, and $Vul$ is updated to the new constraint for the reduced edge set $\{e_1, e_2, \ldots, e_{h-1}\}$ according to Equation No. \ref{eqn13} in Line No. 27. 

\begin{equation} \label{recc}
\scriptsize
dp[h][Vul][0] = \left\{ 
  \begin{array}{ c l }
  \min_{cipher \in cipher\_tab} T_{enc}[v_i, v_j] + T_{dec}[v_i, v_j] & \quad \textrm{if } \text{h = 0} \\
    \min_{cipher \in cipher\_tab} T_{enc}[v_i, v_j] + T_{dec}[v_i, v_j] \\ + dp[h-1][Vul - W_{v_i, v_j} \cdot V_{v_i, v_j}] & \quad \textrm{otherwise}
  \end{array}
\right.
\end{equation}

Note that the weights $W_{v_i, v_j}$ and constraint $UV_{req}$ may be decimals, so we convert them into integers by multiplying with powers of 10.

\begin{algorithm}[!htb]
\caption{Cipher Assignment Algorithm}
	\KwIn{$G(V, E)$, Multi-cloud system parameters, Task to resource mapping ($X$), Security Cipher Table similar to Table \ref{tab3} ($cipher\_tab$), $W_{v_i, v_j}, UV_{req}, UV_{v_i, v_j}$}
	\KwOut{$\mathcal{C}$} 
	Initialize DP table $dp[\lvert E \rvert + 1][UV_{req}+1]$\;
	\For{$h \longleftarrow 1$ to $\lvert E \rvert$}{
		$(v_i, v_j) \longleftarrow e_h$\;
		$VM(k, p) \longleftarrow type(R[X[v_i]]), VM(k', p') \longleftarrow type(R[X[v_j]])$\;
		\For{$Vul \longleftarrow 0$ to $UV_{req}$}{
			$best\_time \longleftarrow INT\_MAX$\;
			$best\_cipher \longleftarrow None$\;
			\For{$cipher \in cipher\_tab$}{
				\If{$W_{v_i, v_j} \cdot V_{cipher} \leq Vul, V_{cipher} \leq UV_{v_i, 						v_j}$} {
						Calculate $T_{enc}[v_i, v_j], T_{dec}[v_i, v_j]$ corresponding to $cipher$ and assuming $v_i, v_j$ are executed on $VM(k, p), VM(k', p')$ respectively from Equation No. 						\ref{eqn14}\;
						\If{$h = 1$} {
							\If{$T_{enc}[v_i, v_j] +T_{dec}[v_i, v_j]  < best\_time$} {
								$best\_time \longleftarrow T_{enc}[v_i, v_j] +T_{dec}[v_i, v_j]$\;
								$best\_cipher \longleftarrow cipher$\;
							}
						}
						\Else {
							$V_{remaining} = Vul - W_{v_i, v_j} \cdot V_{cipher}$\;
							$total\_time \longleftarrow dp[h-1][V_{remaining}][0] + T_{enc}[v_i, v_j] +T_{dec}[v_i, v_j]$\;
							\If{$total\_time < 								best\_time$} {
								$best\_time \longleftarrow total\_time$\;
								$best\_cipher \longleftarrow cipher$\;
							}
						}
				}
			}
			$dp[h][Vul] \longleftarrow [best\_time, best\_cipher]$\;
		}
	}
	$Vul \longleftarrow UV_{req}$\;
	\For{$h \longleftarrow \lvert E \rvert$ to $1$}{
		$(v_i, v_j) \longleftarrow e_h$\;
		$C_{v_i, v_j} \longleftarrow dp[h][Vul][1]$\;
		Add tuple $(v_i, v_j, C_{v_i, v_j})$ to $\mathcal{C}$\;
		$Vul \longleftarrow Vul - W_{v_i, v_j} \cdot V_{v_i, v_j}$\;
	}
	
	\KwRet $\mathcal{C}$\;
	\label{Algo:3}
\end{algorithm}

\subsection{VM Allocation}
We follow an efficient list-based approach to assign resources to tasks. List-based scheduling generally works in two phases: Ordering the tasks based on some rank followed by assigning the resources to the tasks in order one by one. Firstly, we reduce the size of the resource pool under consideration similar to \cite{rodriguez2014deadline} as follows: Let $\mathcal{P}$ denote the largest set of tasks that can be executed in parallel. Then $R$ can be redefined as $R = \{vm_r : r \in [\lvert \mathcal{P} \rvert \cdot n]\}$, by considering a VM instance of each type for each task in $\mathcal{P}$. One way to find the set $\mathcal{P}$ is to find the largest set of tasks, all having the same topological level. Since all the tasks have the same topological level, they have no dependency relation. The set $\mathcal{P}$ computed in this manner may not necessarily be the largest one. Secondly, we define the notion of rank of a task in Definition \ref{def6}.

\begin{definition} \label{def6}
$Rank(v_i)$ denotes the rank of the task $v_i$, which gives an idea of the worst-case processing time along all paths from $v_i$ to $v_n$ formulated in Equation No. \ref{eqn18}.
\end{definition}  

\begin{dmath} \label{eqn18}
\scriptsize
Rank_{v_i} = \left\{ 
  \begin{array}{ c l }
  \max_{v_j \in succ(v_i)}\{Rank(v_j)\} + \overline{T_{exec}[v_i, VM(k, p)]} \\ + \sum_{v_j \in succ(v_i)} \overline{T_{comm}[v_i, v_j]} & \quad \textrm{if } v_i = v_1 \\ \\
  \sum_{v_j \in pred(v_i)} \overline{T_{exec}[v_i, VM(k,p)]} & \quad \textrm{else if } v_i = v_n \\ \\
  \max_{v_j \in succ(v_i)}\{Rank(v_j)\} + \overline{T_{exec}[v_i, VM(k,p)]} \\ + \sum_{v_j \in succ(v_i)} + \overline{T_{comm}[v_i, v_j]} & \quad \textrm{otherwise}
  \end{array}
\right.
\end{dmath}
where $\overline{T_{exec}[v_i, VM(k, p)]}$ denotes the average execution time of executing $v_i$ on VM types, $\overline{T_{comm}[v_i, v_j]} = \frac{w(v_i, v_j)}{\overline{B}}$, $\overline{B}$ denoting the average bandwidth. The allocation algorithm known as List Based Scheduler (LBS) is presented in Algorithm 4 and described as follows: We start allocating tasks bottom-up because the processing time of a task depends majorly on the successor nodes, as seen from Equation No. \ref{eqn4}. For this, we sort the tasks in decreasing order of their topological levels, prioritizing tasks with a higher rank in Line No. 1, 2. Since tasks with the same topological level can execute simultaneously, we assign different VM instances to each of them, and the set $vms\_allocated$ keeps track of the VMs allocated till now, which is initialized to empty set in Line No. 8. We use the $process\_task$ routine to come up with the necessary components in the processing time, denoted by $PT_{v_i, vm_r}$ when $v_i$ is assumed to be executed on $vm_r$ in Line No. 12, 13. Since we know where each successor is allocated, we can compute the $transfer\_time, transfer\_cost, rel$. Since we do not know the cipher assignment yet, we assume $enc\_time, dec\_time$ to be zero. The cost and reliability associated with processing time and data-transfer are calculated in Line No. 14-17. After this, we calculate a metric $metric_{v_i, vm_r}$ based on a linear combination of cost, processing time, and reliability with each parameter normalized using min-max normalization in Line No. 20. The weights $\alpha$, $\beta$, and $\gamma$ give the relative importance between the parameters. In this study, we give more importance to cost and choose the weights: $\alpha = 0.7, \beta = 0.2, \gamma = 0.1$. A VM instance with a lesser value of $metric_{v_i, vm_r}$ is desirable. Hence we sort the instances in increasing order of $metric_{v_i, vm_r}$ and assign the first VM instance yet not assigned to any previous task with the same topological level in Line No. 22-27 to increase parallelizability. Our algorithm differs from list-based methods used in other algorithms \cite{tang2021reliability} \cite{xie2017energy}, in that it allocates tasks in a reverse order, i.e., it allocates a successor task before its predecessor task. This is because, as mentioned earlier, the processing time is mostly dependent on the successor nodes. \\

Lastly, we propose an iterative local search algorithm (LS) that takes as input a VM allocation and reassigns VMs to tasks for better makespan, cost, and reliability, as illustrated in Algorithm 5. The tasks are first ordered by their Ranks in Line No. 1. For a fixed number of iterations, the algorithm begins by reassigning a VM instance $vm_r$ to each task $v_i$ and computing the corresponding $makespan (s_{v_i, vm_r}), cost (cost_{v_i, vm_r}), reliability (\mathcal{R}_{v_i, vm_r})$. Similar to Algorithm 4, we use min-max normalization to come up with a metric with $\alpha = 0.6, \beta = 0.2, \gamma = 0.2$ in Line No. 9. The VM instance $vm_{u_i}$ with least value is assigned to the task $v_i$ in Line No. 10. If there is no change in the allocation, we terminate the procedure as shown in Line No. 11, 12.

\begin{algorithm}[!htb]
	\caption{LBS Allocation Algorithm}
	\KwIn{$G(V, E)$, Multi-cloud system parameters, Resource Pool ($R$)}
    \KwOut{Task to resource mapping ($X$)}
    Sort tasks in decreasing order of their Ranks in list $Q$\;
    Sort tasks in decreasing order of their topological levels, break ties based on the order in $Q$. Store the resulting order in list $L$\;
    $prev\_lvl \longleftarrow -1$\;
    $vms\_allocated \longleftarrow \emptyset$\;
    \For{$v_i \in L$} {
    	$curr\_lvl \longleftarrow top\_level(v_i)$\;
    	\If{$prev\_lvl \neq curr\_lvl$} {
    		$vms\_allocated \longleftarrow \emptyset$\;
    	}
    	\For{$vm_r \in R$} {
        	$VM(k, p) \longleftarrow type(vm_r)$\;
        	$exec\_time \longleftarrow \frac{w(v_i)}{w(VM(k, p))}$\;
    		$dec\_time, transfer\_time, transfer\_cost,$ $enc\_time, rel \longleftarrow process\_task(v_i)$\;
    		$PT_{v_i, vm_r} \longleftarrow dec\_time + exec\_time + transfer\_time + enc\_time$\;
    		Calculate $lease\_cost$ corresponding to lease period $PT_{v_i, vm_r}$ using Equation No. \ref{eqn8}, \ref{eqn9}\;
    		$cost_{v_i, vm_r} \longleftarrow lease\_cost + transfer\_cost$\;
    		 Calculate $lease\_rel$ corresponding to lease period $PT_{v_i, vm_r}$ using Equation No. \ref{eqn16}\;
    		 $\mathcal{R}_{v_i, vm_r} \longleftarrow rel \cdot lease\_rel$\;
    	}
    	Calculate $PT_{v_i, \min}, PT_{v_i, \max}, cost_{v_i, \min}, cost_{v_i, \max},$ \\ $\mathcal{R}_{v_i, \min}, \mathcal{R}_{v_i, \max}$ over all VM instances in $R$\;
    	\For{$vm_r \in R$} {
    		$metric_{v_i, vm_r} \longleftarrow \alpha \cdot \frac{cost_{v_i, vm_r} - cost_{v_i, \min}}{cost_{v_i, \max} - cost_{v_i, \min}} + \beta \cdot \frac{PT_{v_i, vm_r} - PT_{v_i, \min}}{PT_{v_i, \max} - PT_{v_i, \min}}   + \gamma \cdot \frac{\mathcal{R}_{v_i, \max} - \mathcal{R}_{v_i, vm_r}}{\mathcal{R}_{v_i, \max} - \mathcal{R}_{v_i, \min}}$
    	}
    	Sort the VM instances in increasing order of $metric_{v_i, vm_r}$ in list $U$\;
    	\For{$vm_r \in U$} {
    		\If{$vm_r \notin vms\_allocated$} {
    			$X[v_i] \longleftarrow r$\;
    			$vms\_allocated \longleftarrow vms\_allocated \cup \{vm_r\}$\;
    			\textbf{break}\;
    		}
    	}
    	$prev\_lvl \longleftarrow curr\_lvl$\;
    }
	\KwRet $X$\;
	\label{Algo:4}
\end{algorithm}
\begin{algorithm}[!htb]
	\caption{LS Algorithm}
	\KwIn{$G(V, E)$, Multi-cloud system parameters, Resource Pool ($R$), Task to resource mapping ($X$)}
    \KwOut{New task to resource mapping ($X'$)}
    Sort tasks in decreasing order of their Ranks in list $L$\;
    \For{$itr \longleftarrow 1$ to $num\_iter$} {
    	\For{$v_i \in L$} {
    		\For{$vm_r \in R$} {
    			Assign $vm_r$ to $v_i$ to get new schedule $X'$\;
    			Calculate $s_{v_i, vm_r}, cost_{v_i, vm_r}, \mathcal{R}_{v_i, vm_r}$ corresponding to $X'$ using Algorithm 1\;
        	}
        	Calculate $s_{v_i, \min}, s_{v_i, \max}, cost_{v_i, \min}, cost_{v_i, \max}$ \\ $\mathcal{R}_{v_i, \min}, \mathcal{R}_{v_i, \max}$ over all VM instances in $R$.\;
    	$vm_{u_i} \longleftarrow \min_{vm_r \in R} \{ \alpha \cdot \frac{cost_{v_i, vm_r} - cost_{v_i, \min}}{cost_{v_i, \max - cost_{v_i, \min}}} + \beta \cdot \frac{s_{v_i, vm_r} - s_{v_i, \min}}{s_{v_i, \max} - s_{v_i, \min}} + \gamma \cdot \frac{\mathcal{R}_{v_i, \max} - \mathcal{R}_{v_i, vm_r}}{\mathcal{R}_{v_i, \max} - \mathcal{R}_{v_i, \min}} \}$\;
    	$X'[v_i] \longleftarrow u_i$\;
    }
    \If{$X' = X$} {
    	\textbf{break}\;
    }
    }
	\KwRet $X'$\;
	\label{Algo:5}
\end{algorithm}
The overall algorithm proceeds by the below-listed steps:
\begin{itemize}
	\item Calculate the reduced resource pool $R$ using BFS \cite{bundy1984breadth}.
	\item Use Algorithm 4 to obtain the task to VM allocation using Algorithm 4.
	\item (Optional) Use Algorithm 5 to get an improved task for VM allocation.
	\item Compute the edge to cipher mapping $\mathcal{C}$ corresponding to the obtained task to VM allocation using Algorithm 3.
	\item Use Algorithm 1 to compute the final schedule $\mathcal{S}$.
	
\end{itemize}

\subsection{Complexity Analysis}
The time complexity for the $process\_task$ routine is $\mathcal{O}(n)$ because each task can have at most $n$ successors or predecessors, hence the time needed for one task in Algorithm 1 is $\mathcal{O}(n)$. For all tasks, the total time needed will be $\mathcal{O}(n^2)$. Finally iterating through $R_{curr}$ in Line No. 25 of Algorithm 1 takes at most $\mathcal{O}(\mathcal{N} \cdot n)$ time as $\lvert R_{curr} \rvert \leq \mathcal{N} \cdot n$. Algorithm 3 is a DP-based approach, and the time needed for computing one entry $dp[i][j]$ depends on the number of ciphers available for encryption $\lvert cipher\_tab \rvert$ as seen from Line No. 7. Hence, the total time needed to compute all entries will be $\mathcal{O}(\lvert cipher\_tab \rvert \cdot UV_{req} \cdot n^2)$ as $\lvert E \rvert \leq n^2$. In Algorithm 4, computing the topological levels and Ranks takes $\mathcal{O}(n^2)$ time. For each task, for each VM instance pair, calculating the processing time, cost and reliability take $\mathcal{O}(n)$ time as seen from the $process\_task$ routine. Sorting the metrics over all VM instance pairs takes $\mathcal{O}(\lvert R \rvert \cdot log \lvert R \rvert)$ time. Hence for one task, the time required is $\mathcal{O}(\lvert R \rvert \cdot (n + \log \lvert R \rvert))$. For all tasks, the time complexity will be $\mathcal{O}(n \cdot \lvert R \rvert \cdot (n + \log \lvert R \rvert))$. As $\lvert R \rvert \leq \mathcal{N} \cdot n$, the time complexity will be at most $\mathcal{O}(\mathcal{N} \cdot n^2 \cdot (n + \log n + \log \lvert \mathcal{N} \rvert))$. Most approaches use evolutionary methods for VM allocation \cite{choudhary2018gsa}, \cite{beegom2015genetic}, \cite{guo2018cost}, which are very time-consuming compared to Algorithm 4. Lastly, we look at the time needed for Algorithm 5. For each reassignment of a VM instance to a task in Line No. 5, we compute the schedule in Line No. 6, which takes $\mathcal{O}(n^2)$ time. Hence the total time needed over all iterations, tasks, and VM instances will be $\mathcal{O}(num\_iter \cdot \mathcal{N} \cdot n^4)$.
\section{Experimental Evaluation} \label{Sec:Experiments}
In this section, we describe the experimental evaluation of the proposed solution approach. First, we start by describing the experimental setup and example task graphs.
\subsection{Experimental setup, Task graphs, and Algorithms}
We implement the proposed methodology on a workbench system with i7 $12^{\text{th}}$ generation processor and 16GB memory in Python 3.8.10. For this study, we consider two real-world task graphs: Epigenomics and LIGO, which are widely used in literature \cite{tang2021reliability}, \cite{szabo2012evolving} for comparison. We consider Epigenomics workflows with $n = 24, 100$ and Cybershake workflows with $n = 30, 100$ \cite{szabo2012evolving}. The structure of each workflow can be obtained in XML format from the website \cite{graphgen}. The smaller size of each workflow is called Small (S), and the larger one is called Large (L). Due to space limitation, we only consider two workflows with two sizes each. More information about the workflows can be obtained from \cite{bharathi2008characterization}.

\subsection{Competitive Algorithms}
\par LBS is compared with the following methods:
\begin{itemize}
	\item Gravitational Search Algorithm (GSA) \cite{choudhary2018gsa}: It aims to minimize the makespan and cost of scheduling the workflow in single cloud. 
	\item Fault-tolerant Cost-Efficient workflow Scheduling (FCWS) \cite{tang2021reliability}: This is a list-based approach for scheduling workflows on multi-cloud systems taking into account the makespan, cost and reliability. VMs are assigned based on a linear combination of cost and reliability. Fault-tolerance is based on hazard rate of the distribution used for reliability analysis. In our case of Poisson distribution, the condition for fault-tolerance used in FCWS algorithm does not apply, so we do not consider it. Also FCWS does not consider communication cost while assigning VMs, so we include it in the cost while assigning VMs.  
\end{itemize}
Since LBS, GSA and FCWS are task to VM allocation algorithms, we use Algorithm 3 on top of them to decide the ciphers for encryption. We do not compare with other workflow scheduling algorithms that take into account security constraints like \cite{wen2016cost}, because it does not consider makespan.

\subsection{Experimental description}
We implement the proposed methodology on a workstation having $i5$, $10^{\text{th}}$ generation processor and 32GB memory in Python 3.8.10. Parameters of our multi-cloud system are set as in \cite{tang2021reliability}. VMs are set with varying computation capacities from 1 to 32. The average bandwidth internal to a cloud is set to 20Mbps, and external bandwidth is set to 100Mbps. Pricing mechanisms are set proportional to VM compute capacities as in Table \ref{tab1} and data transfer costs according to Table \ref{tab2}. The boot time of each VM was set to 97 seconds \cite{rodriguez2014deadline}. The number of cloud providers considered is six, two from each type of \emph{MA}, \emph{AWS}, and \emph{GCP}. Two cloud providers with the same types are assumed to be located in different centers. The encryption levels and overhead is chosen as in Table \ref{tab3}. We assume that $w_{k, p} =1$. The security constraints are set as follows \cite{jiang2017design}: $UV_{req} = \eta \cdot V_{\max}$, where $\eta \in [0.1, 0.7]$ in steps of 0.1. $UV_{v_i, v_j}$ is a randomly chosen security cipher from Table \ref{tab3} and the weights $W_{v_i, v_j} \in [0.1, 1]$. The parameter of the Poisson Distribution is set uniformly at random $\lambda \in [10^{-8}, 10^{-7}]$. We use Algorithm 5 on top of each algorithm to see the improvement obtained. Each experiment is performed 15 times, and the results are reported on average.

\begin{figure*}[!ht]
\centering
\begin{tabular}{ccc}
\includegraphics[width=0.35\textwidth]{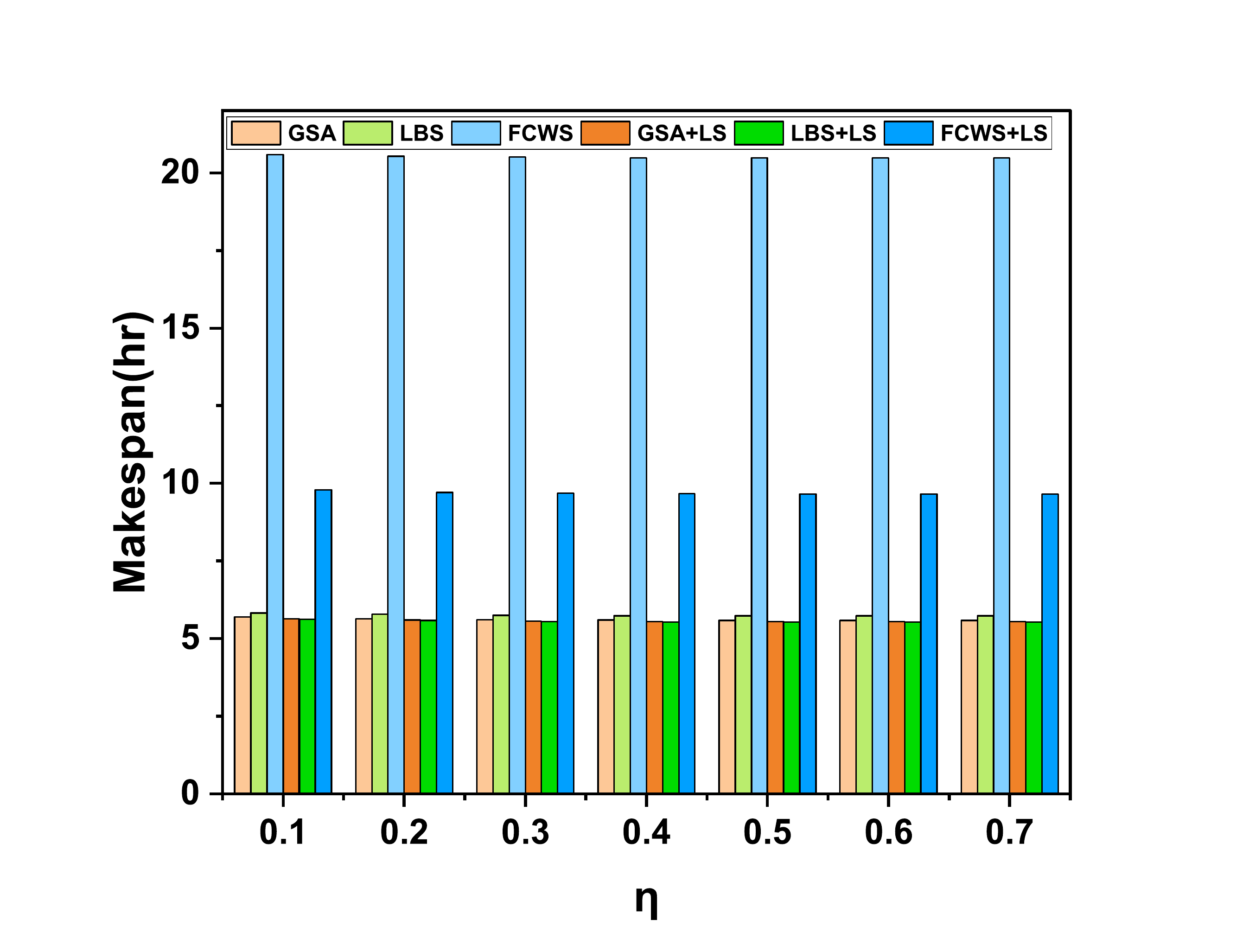} &
\includegraphics[width=0.35\textwidth]{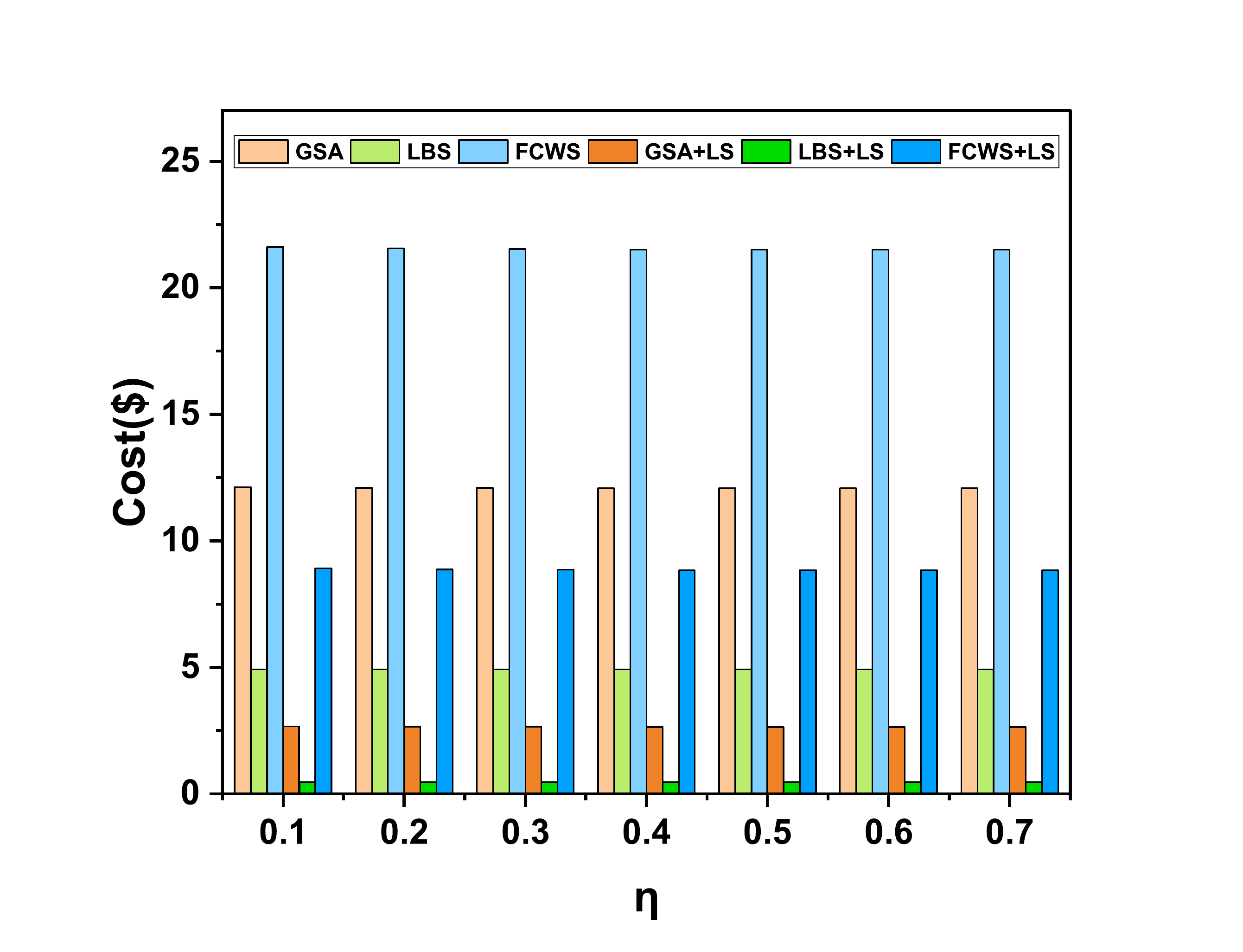} &
\includegraphics[width=0.35\textwidth]{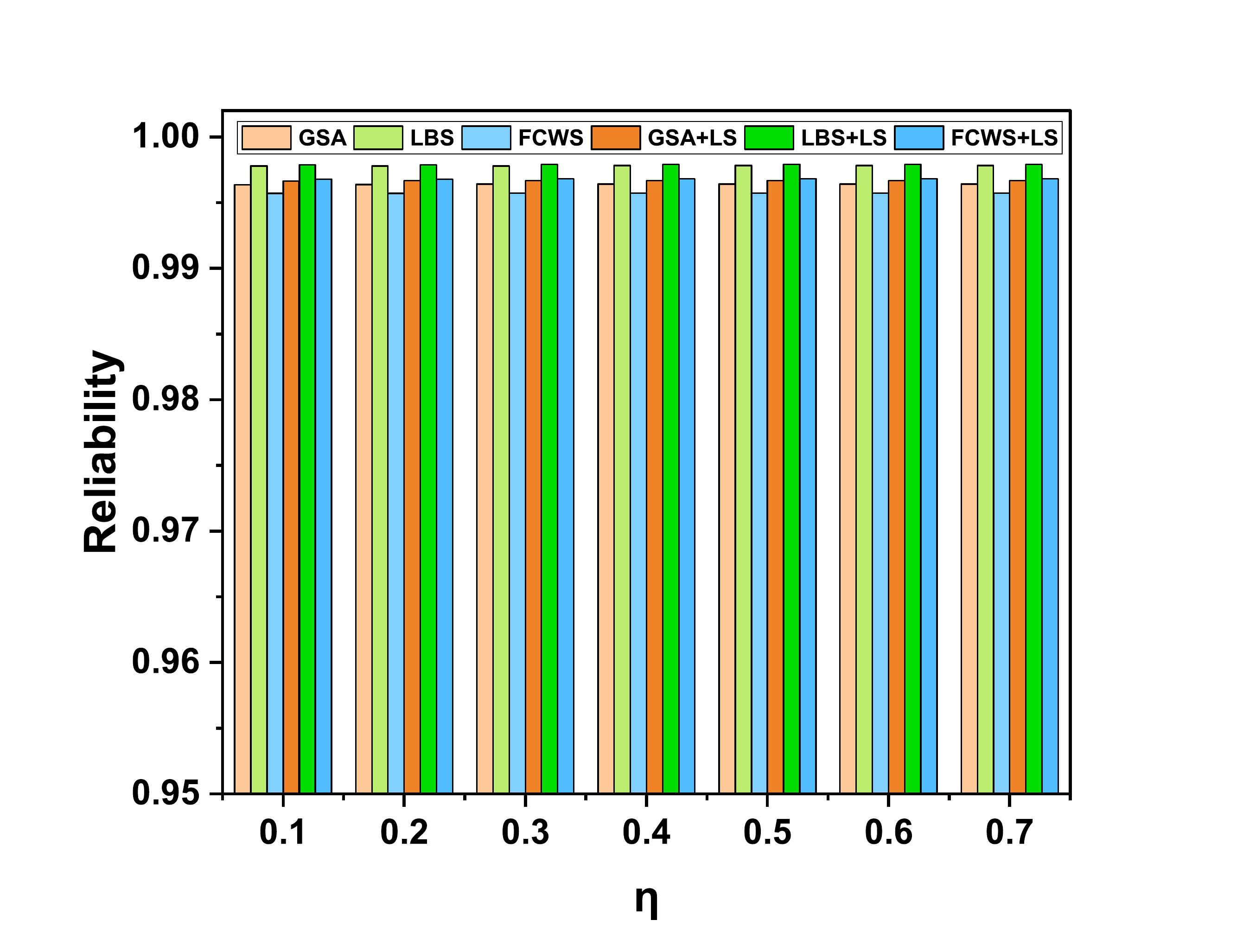}
\\
(a) & (b) & (c) \\
\end{tabular}
\caption{Epigenomics graph with $n = 24$}
\label{Fig1}
\end{figure*}

\begin{figure*}[!ht]
\centering
\begin{tabular}{ccc}
\includegraphics[width=0.35\textwidth]{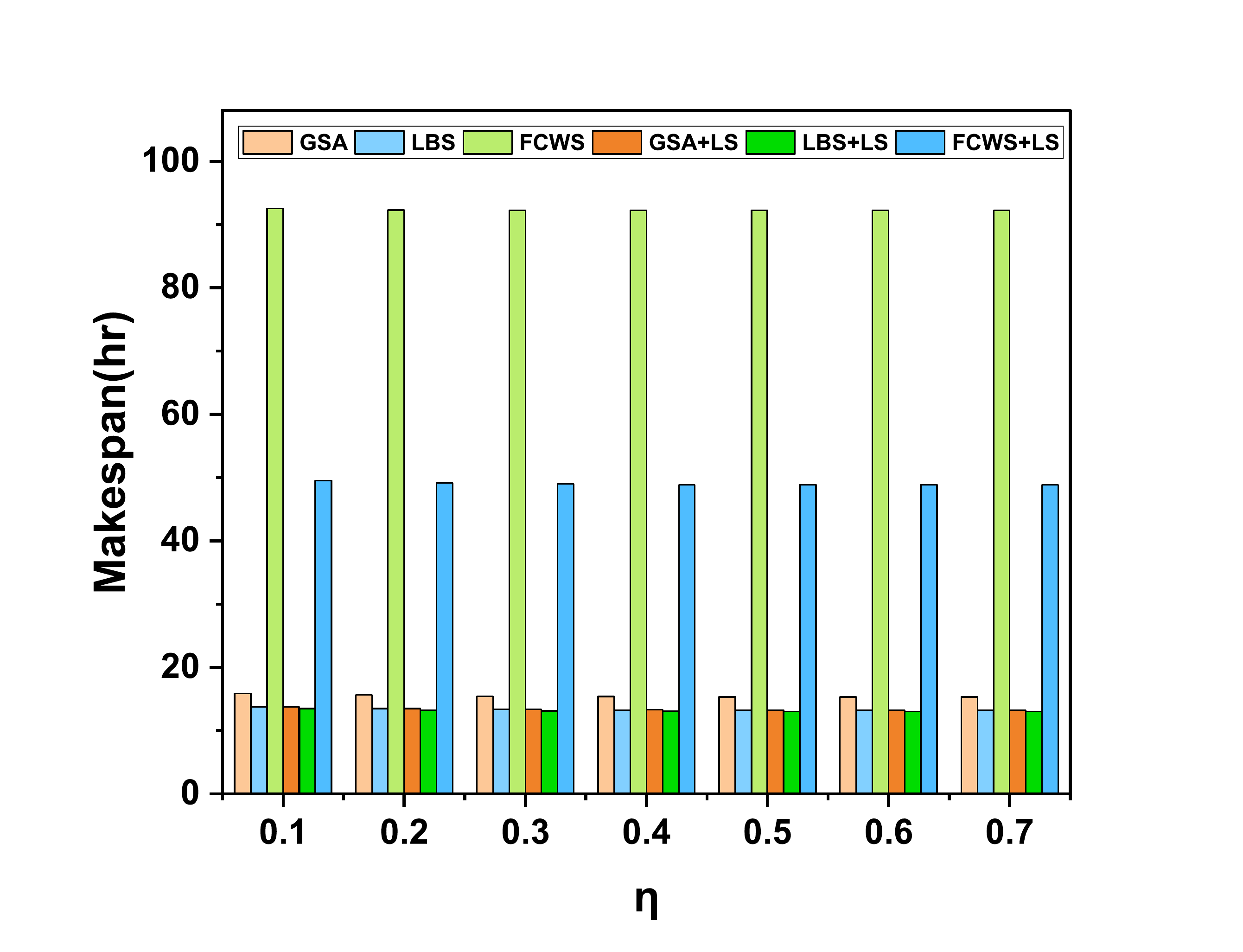} &
\includegraphics[width=0.35\textwidth]{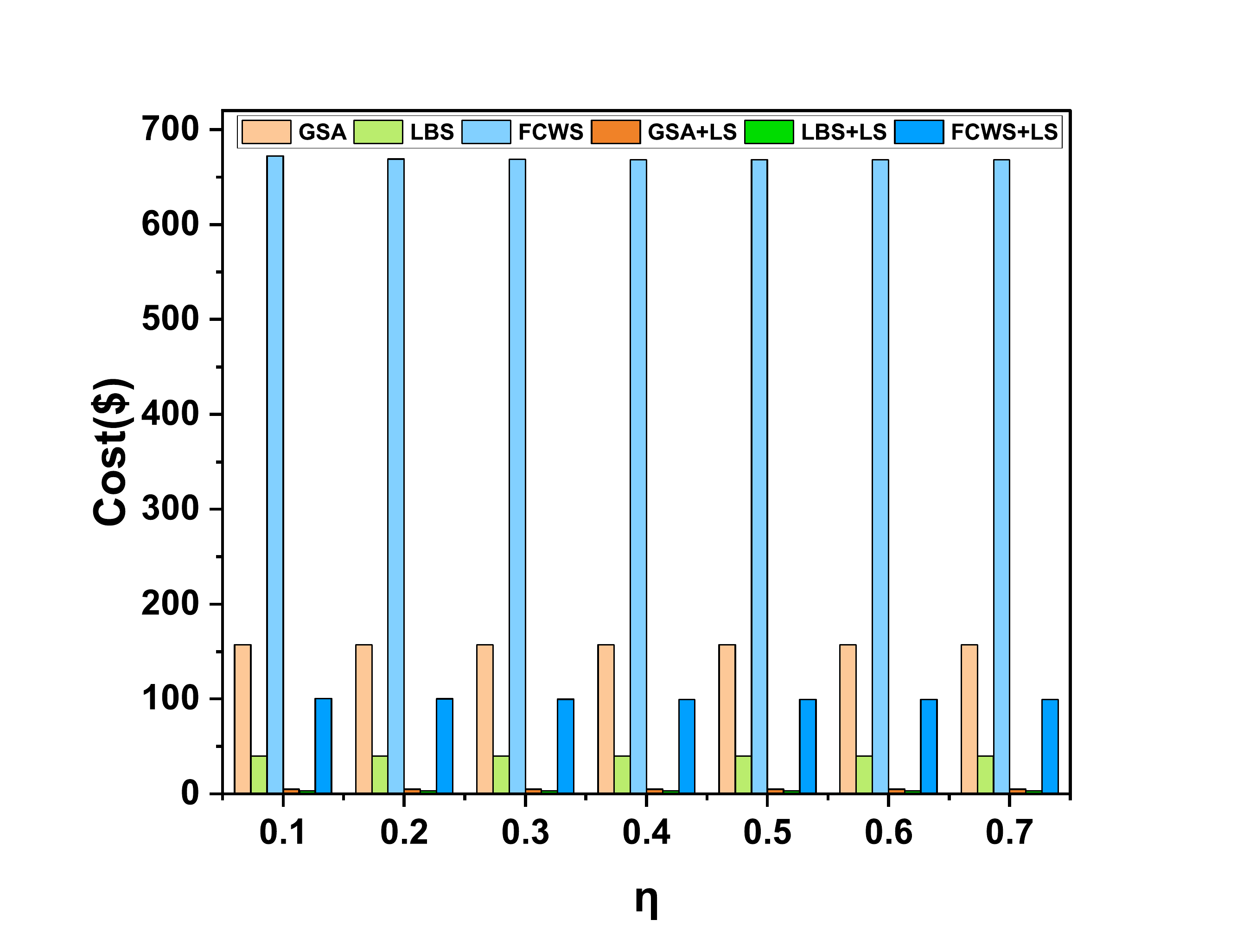} &
\includegraphics[width=0.35\textwidth]{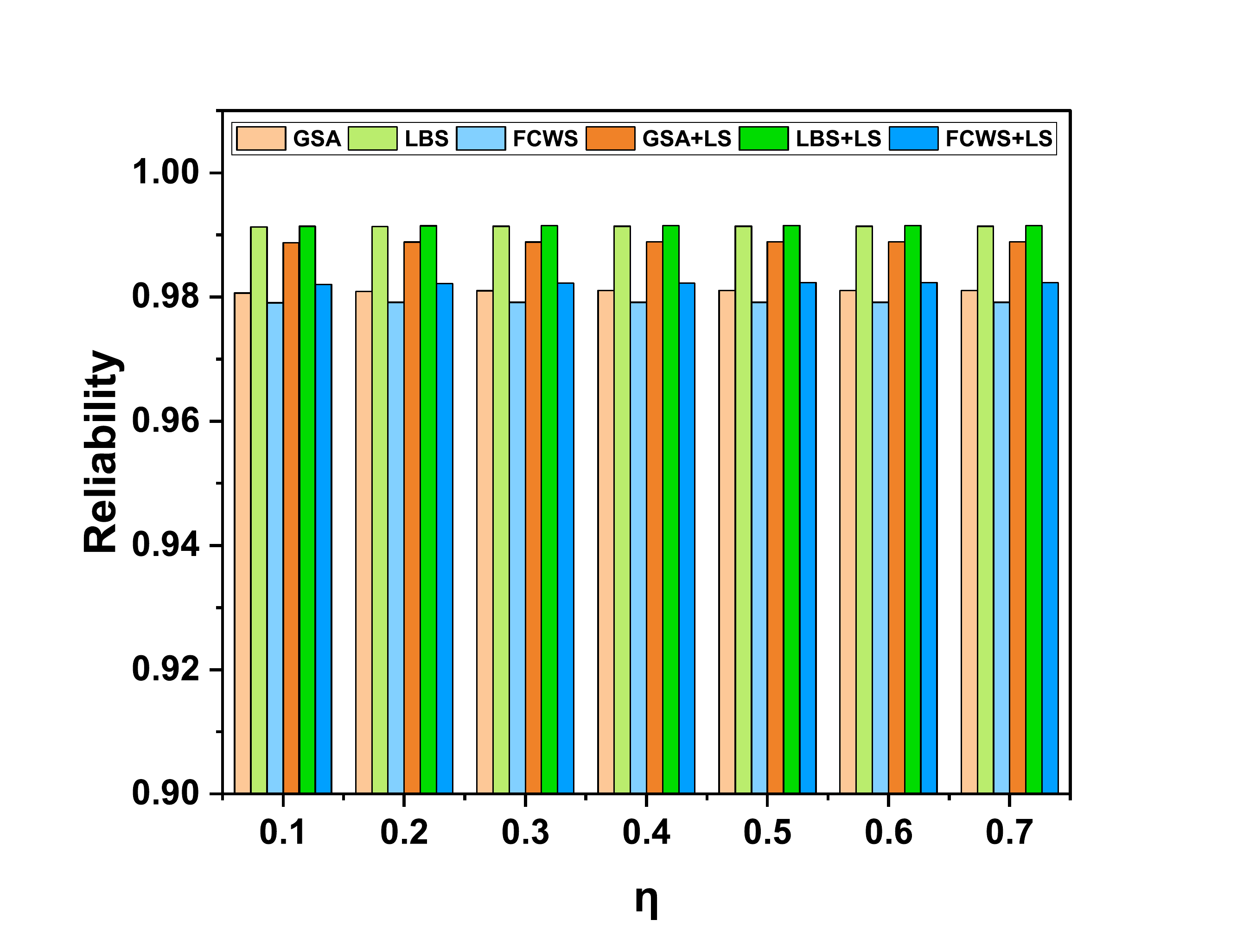} 
\\
(a) & (b) & (c) \\
\end{tabular}
\caption{Epigenomics graph with $n = 100$}
\label{Fig2}
\end{figure*}

\begin{figure*}[!ht]
\centering
\begin{tabular}{ccc}
\includegraphics[width=0.35\textwidth]{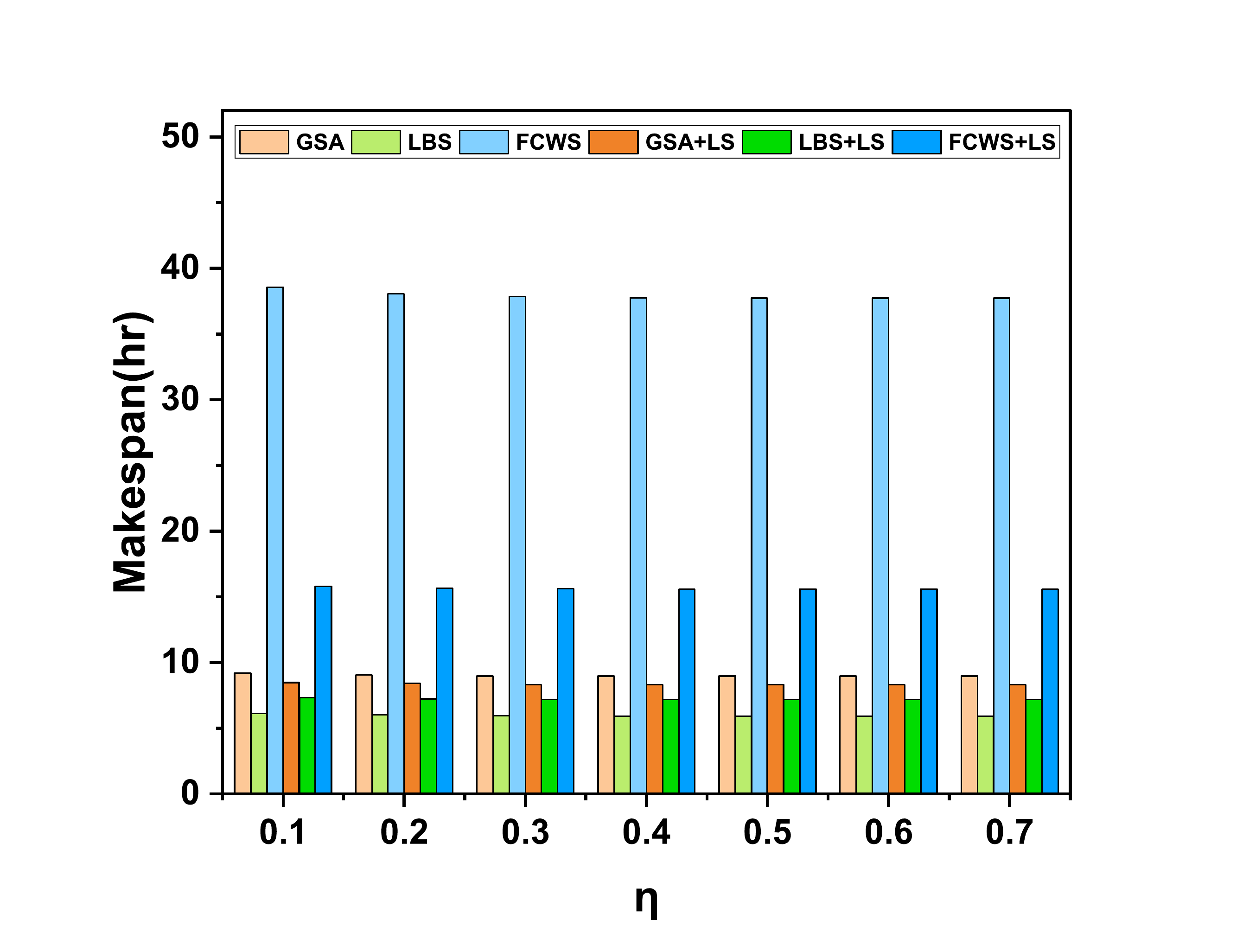} &
\includegraphics[width=0.35\textwidth]{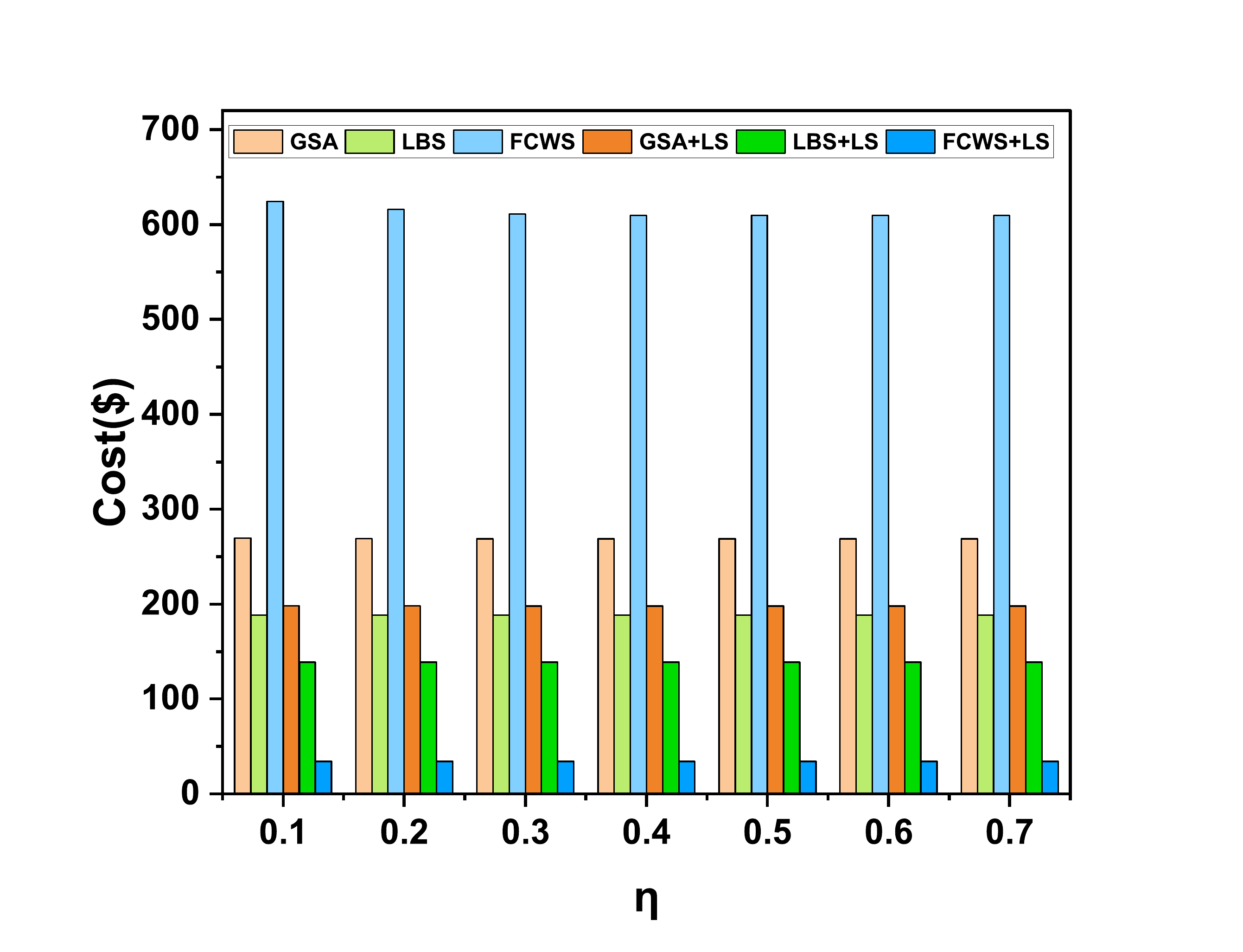} &
\includegraphics[width=0.35\textwidth]{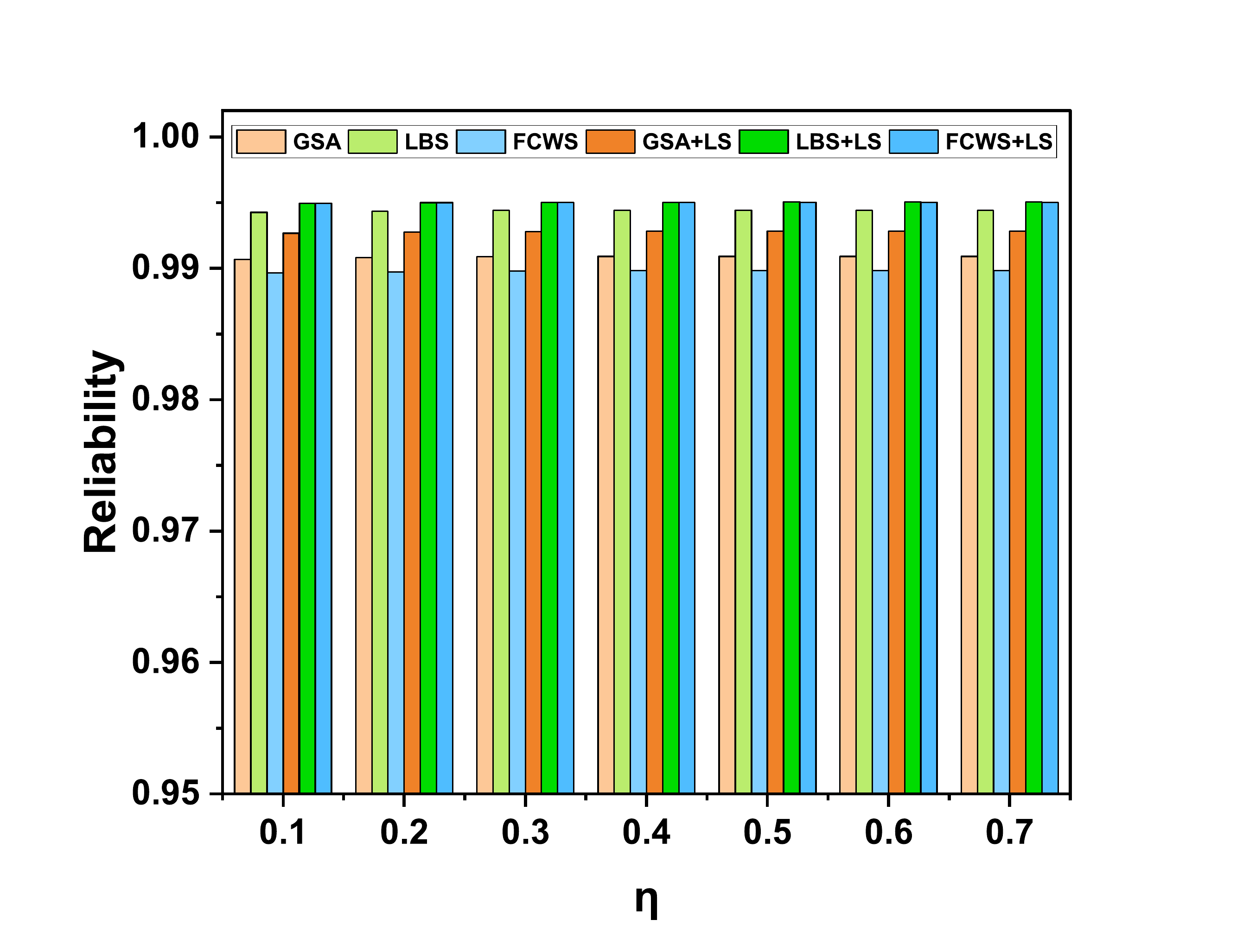}
\\
(a) & (b) & (c) \\
\end{tabular}
\caption{Cybershake graph with $n = 30$}
\label{Fig3}
\end{figure*}

\begin{figure*}[!ht]
\centering
\begin{tabular}{ccc}
\includegraphics[width=0.35\textwidth]{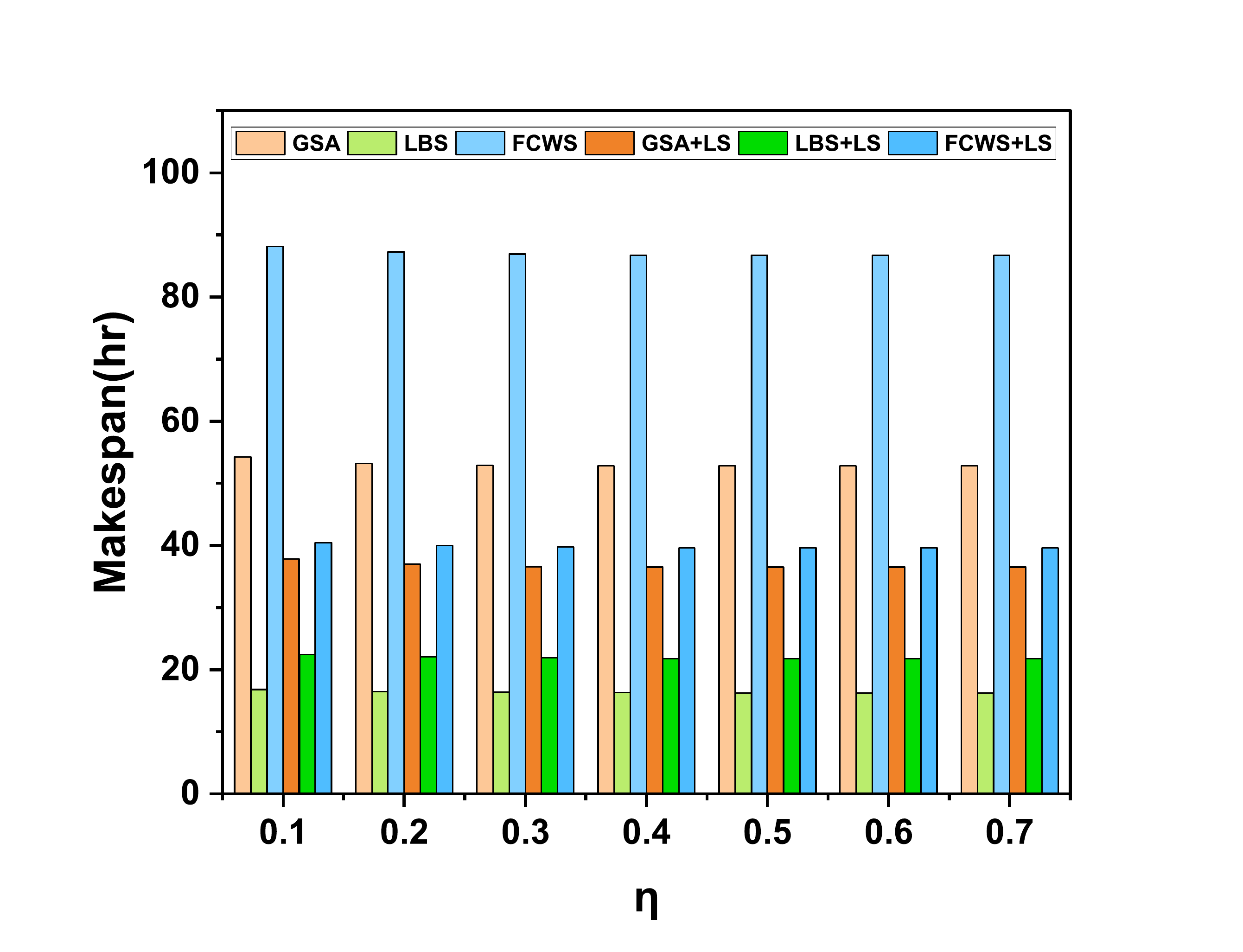} &
\includegraphics[width=0.35\textwidth]{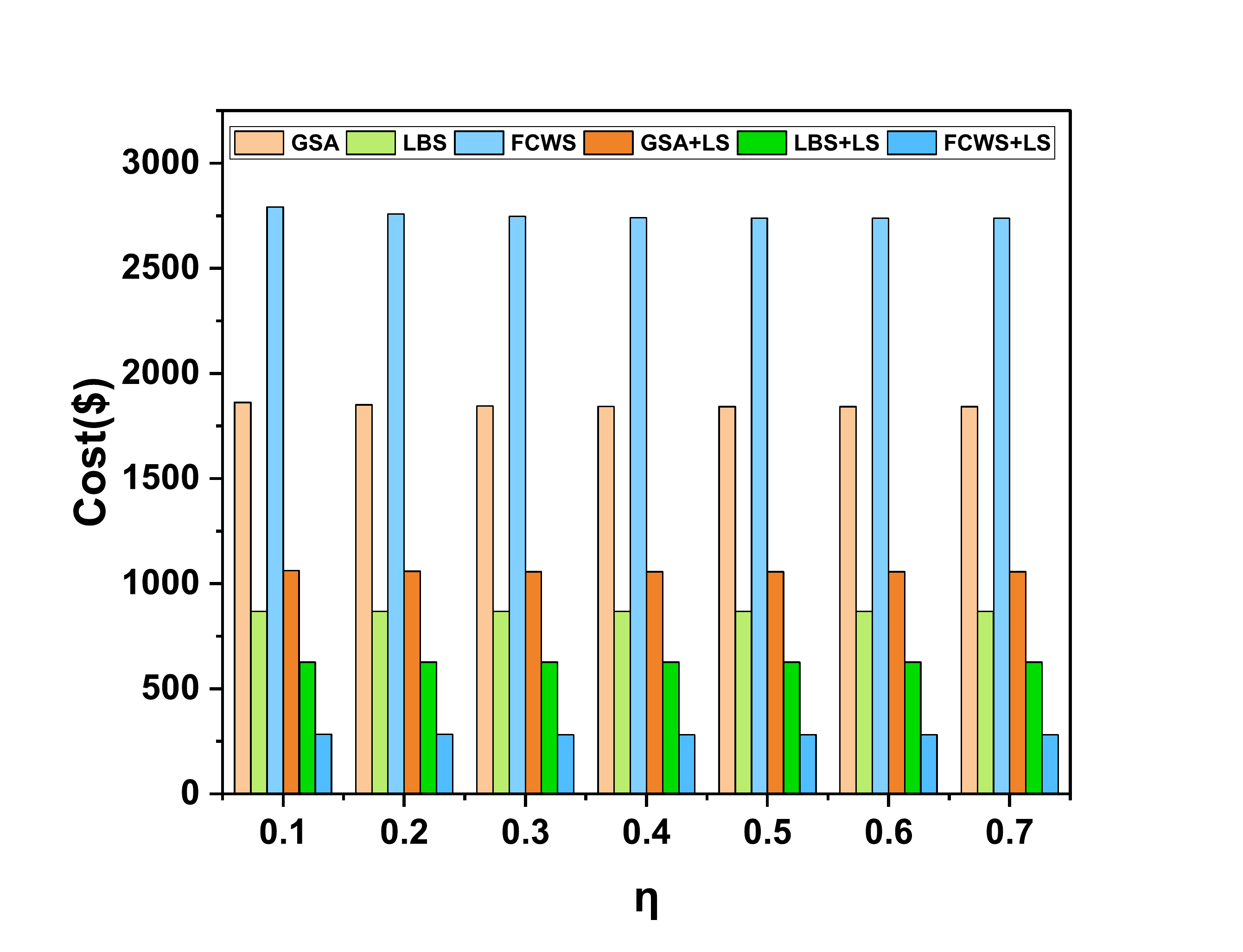} &
\includegraphics[width=0.35\textwidth]{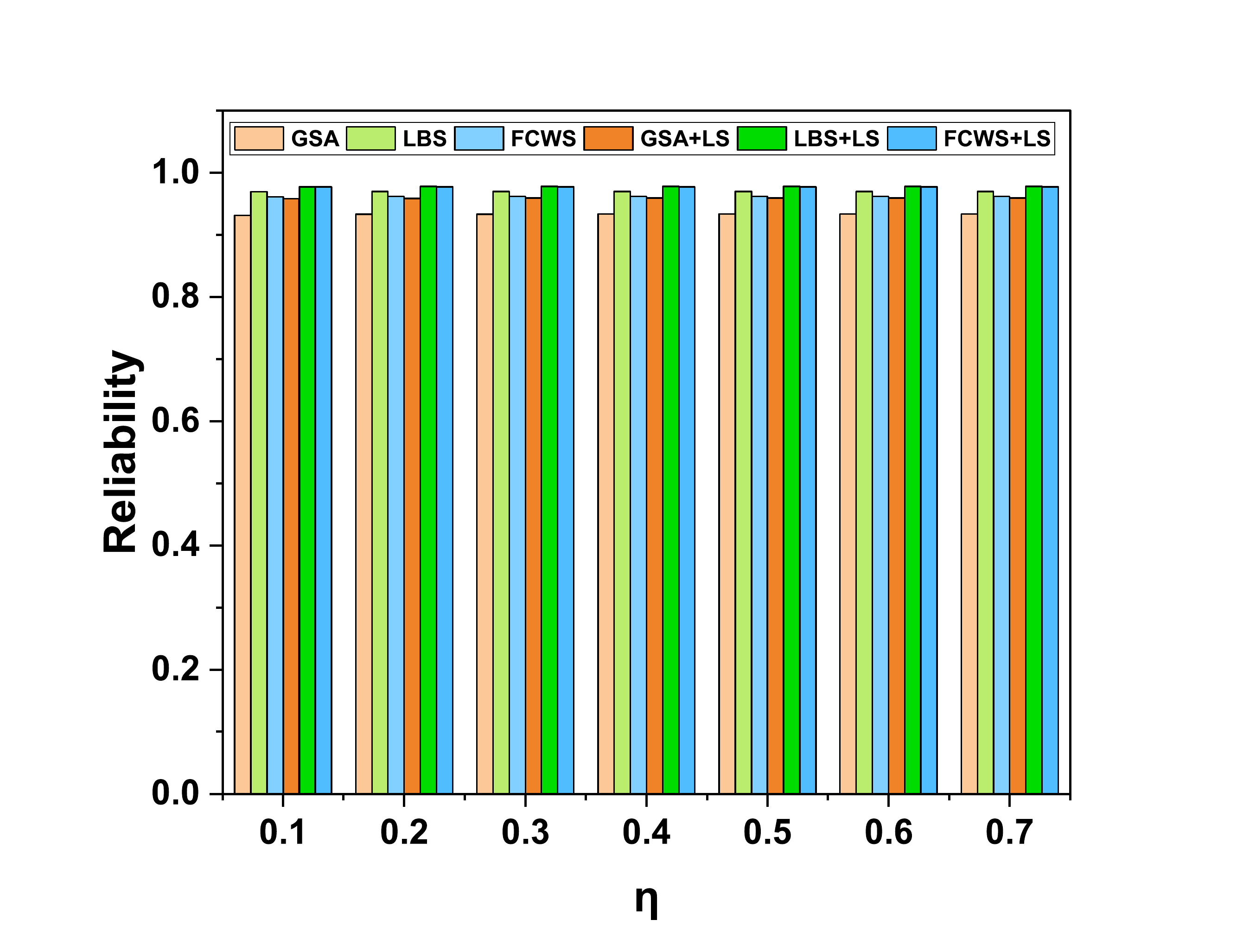}
\\
(a) & (b) & (c) \\
\end{tabular}
\caption{Cybershake graph with $n = 100$}
\label{Fig4}
\end{figure*}

\begin{enumerate}
\item \emph{Epigenomics:}
For Small size, GSA gives the least makespan with LBS having slightly more makespan, both perform much better than FCWS by 71.93\% as seen from Fig. \ref{Fig1} (a). From Fig. \ref{Fig1} (b) LBS gives much lesser cost than GSA by 59.37\% followed by FCWS by 77.18\%. From Fig. \ref{Fig1} (c), in terms of reliability, all algorithms perform similarly with LBS performing slightly better than GSA by 0.15\% and FCWS by 0.2\%. The application of LS gives a slight improvement in makespan for GSA and LBS and a significant improvement in FCWS by 52.74\%. The improvement in cost is significant in all algorithms: 78.09\% for GSA, 90.52\% for LBS, and 58.86\% for FCWS. The improvement in reliability is not so much.
\par For Large size, LBS gives the least makespan outperforming GSA by 13.6\% followed by FCWS by 85.5\% from Fig. \ref{Fig2} (a). From Fig. \ref{Fig2} (b) in terms of cost, LBS performs significantly better than GSA by 74.66\% and FCWS by 94.04\%. In terms of reliability, GSA and FCWS perform similarly and LBS outperforms them by 1.06\% as seen from Fig \ref{Fig2} (c). The application of LS gives a slight improvement in makespan for LBS and significant improvement in GSA by 13.47\% and  FCWS by 52.74\%. The improvement in cost is significant in all algorithms: 96.84\% for GSA, 92.02\% for LBS, and 85.09\% for FCWS. 

\item \emph{Cybershake:} For Small size, LBS gives the least makespan outperforming GSA by 33.81\% followed by FCWS by 84.27\% as seen in Fig. \ref{Fig3} (a). From Fig \ref{Fig3} (b) LBS gives the least cost outperforming GSA by 29.92\% and FCWS by 69.24\%. From Fig. \ref{Fig3} (c), in terms of reliability, all algorithms perform similarly. The application of LS gives a degraded makespan of LBS by 20.71\% while giving a significant improvement in GSA by 7.34\% and FCWS by 58.78\%. The improvement in cost is significant in all algorithms: 26.39\% for GSA, 26.24\% for LBS, and 94.44\% for FCWS. The improvement in reliability is considerable only in GSA by 0.53\%.
\par For Large size, LBS gives the least makespan outperforming GSA by 69.12\% followed by FCWS by 81.16\% from Fig. \ref{Fig2} (a). From Fig. \ref{Fig2} (b) in terms of cost, LBS performs significantly better than GSA by 53.02\% and FCWS by 68.47\%. In terms of reliability, LBS gives the most reliability outperforming FCWS by 0.84\% and GSA by 3.92\% as seen from Fig \ref{Fig2} (c). The application of LS gives a degraded makespan for LBS by 33.83\% while giving a significant improvement in GSA by 30.72\% and  FCWS by 54.26\%. The improvement in cost, and reliability is significant in all algorithms: 42.69\%, 2.77\% for GSA, 27.73\%, 0.84\% for LBS, and 89.75\%, 1.63\% for FCWS. 
 
\end{enumerate}
In summary, as the task graph size increases, performance degrades because we have more task nodes and hence more processing time is needed. As the value of $\eta$ increases, makespan,  and cost decrease while reliability increases. This is because when the security constraint becomes looser, we can use ciphers with less overhead to encrypt data reducing the processing time. LBS always outperforms the other algorithms in terms of cost and reliability and makespan for the majority of the cases. The application of LS brings improvement in  makespan, cost, and reliability of GSA and FCWS. For LBS, the application of LS always gives improvements in cost and reliability. The makespan also improves in some cases.

\section{Conclusion and Future Research Directions}\label{Sec:CFA}
In this paper, we have studied the problem of scheduling the tasks present in a scientific workflow in a multi-cloud system. The goal of the scheduling is to minimize the makespan and cost and to maximize reliability subject to security constraints. For this problem, we have developed an efficient solution methodology. This has been validated with real-life scientific workflows. In the future, we would like to include fault tolerance in our model and look for more efficient solution methodologies.
%
%
%



%
%

\bibliographystyle{spbasic}      
\bibliography{Paper}   


\end{document}